\documentclass[12pt, letterpaper]{article}
\usepackage{harvard,fullpage,lscape,amsmath}
\usepackage{graphicx}
\usepackage{setspace}

\begin{document}

\title{Party Polarization in Congress:\\ A Network Science Approach}

\author{Andrew Scott Waugh, {\it University of California, San Diego}\\
Liuyi Pei, {\it University of California, Irvine}\\
James H. Fowler, {\it University of California, San Diego}\\
Peter J. Mucha, {\it University of North Carolina at Chapel Hill}\\
Mason A. Porter, {\it University of Oxford}
}

\maketitle
\graphicspath{{/work/modularity/modularity_figs/}}

\singlespacing{\it We measure polarization in the United States Congress using the network science concept of \emph{modularity}.  Modularity provides a conceptually-clear measure of polarization that reveals both the number of relevant groups and the strength of inter-group divisions without making restrictive assumptions about the structure of the party system or the shape of legislator utilities.  We show that party influence on Congressional blocs varies widely throughout history, and that existing measures underestimate polarization in periods with weak party structures. We demonstrate that modularity is a significant predictor of changes in majority party and that turnover is more prevalent at medium levels of modularity. We show that two variables related to modularity, called `divisiveness' and `solidarity,' are significant predictors of reelection success for individual House members.  Our results suggest that modularity can serve as an early warning of changing group dynamics, which are reflected only later by changes in party labels.}


\doublespacing

\section{Introduction} \label{sec1}

A great deal of recent research has been devoted to identifying and explaining polarization in the United States.  Indeed, there has been enough to fill the pages of two annual review articles in the past half-decade (Layman et al. 2006, Fiorina \& Abrams 2008)\nocite{layman2006,fiorina2008}.  At its core, the polarization debate began with an observation about the nature of partisanship in Congress.  Starting in the late 1970s, researchers began to notice increases in intra-party cohesion and decreases in inter-party cooperation on roll-call votes (McCarty et al. 2007)\nocite{mccarty2007}.  This finding puzzled scholars because it ran counter to empirical evidence for the weakening of partisanship in the electorate \cite{coleman1996b} and the theoretical expectation that institutional incentives should drive parties to adopt similar (i.e., median) policy positions \cite{downs1957}.  

These puzzles form the fault lines of the two major conflicts in the literature on partisan polarization.  The first conflict attempts to locate the origins of partisan polarization either in the electorate (McCarty et al. 2007)\nocite{mccarty2007}, among political elites \cite{fiorina2008}, or in a combination of the two \cite{jacobson2005,jacobson2006}.  The second attempts to explain the relative influences of party and ideology on Congressional voting, and well-known publications have argued for party effects \cite{cox1993,cox2005,smith2007}, ideological effects \cite{krehbiel1991}, and interactions between the two \cite{rohde1991,aldrich2001}.  In both instances, researchers have pondered how non-median party positions result from institutions that in theory lend themselves towards moderation (Layman et al. 2006)\nocite{layman2006}.

We argue that the polarization debate has been limited by its overemphasis on expectations derived from spatial models of ideology.  By contrast, we define polarization behaviorally---rather than ideologically---based on the voting decisions made by the legislators and the way those decisions divide them into distinct groups.  To operationalize this definition, we use the tools of network science.  Suppose that each legislator is a `node' in the network and that the level of agreement between two legislators in roll-call voting indicates the strength of a `tie' between them.\footnote{As discussed in more detail in the online supplementary information, the ties between all pairs of legislators in one legislative body (Senate or House of Representatives) in a single Congress are described by an \emph{adjacency matrix} ${\bf A}$ with elements
\begin{equation}
	A_{ij} =\frac{1}{b_{ij}}\sum_{k} \gamma_{ijk}\,,
\end{equation}
where $\gamma_{ijk}$ equals $1$ if legislators $i$ and $j$ voted the same on bill $k$ and $0$ otherwise, and $b_{ij}$ is the total number of bills on which both legislators voted. 
}  In a highly-polarized legislature, we reason that groups like parties contain strong ties between individuals within the same group but relatively weak ties between individuals in different them.  In the extreme case of pure party-line voting, all members of the same party vote identically and therefore have the strongest possible ties between each other.  They also have the weakest possible ties to members of the other parties.  In contrast, individuals in a legislature with low polarization tend to have ties both to individuals in their own group and to those in other groups, weakening the strength of group divisions.  Here, we utilize network \emph{modularity} to first identify relevant `communities' (tightly-knit groups) in Congress and then quantify the severity of such a division.

In section \ref{rr_bg}, we use insights from Downs's \citeyear{downs1957} model of ideology to explain our critique of spatial models as tools for measuring partisan polarization.  In section \ref{sec2}, we explain the intuition behind our use of modularity and discuss its advantages over existing measures.  In sections \ref{sec4} and \ref{comcent}, we present empirical evidence for the utility of modularity in predicting majority-party switches Congress as well as the electoral fortunes of individual legislators.  We conclude in section \ref{sec5} by discussing the ongoing role for modularity in advancing our understanding of party polarization.  In online supplementary information, we provide further details of the network-science methodology, additional figures and tables, and a descriptive example from the 19th century.


\section{Ideology, Information, and Polarization} \label{rr_bg}

Political scientists have focused intently on Downs's median voter theorem \cite{downs1957} and of the spatial model at its core.  However, although the median voter result is normatively appealing, it is also rather fragile.  In the Congressional case, adding influential activists \cite{aldrich1995} or rules favoring the majority party \cite{cox1993,cox2005} to the model results in ideological divergence of parties.  In the presidential case, empirical investigations show that parties alter the composition of the electorate by appealing to core supporters, which allows them to maintain extreme positions \cite{holbrook2005}.  These results challenge the expectation of median outcomes while maintaining strong assumptions about the nature of ideology, especially as perceived by the electorate.  We argue that these assumptions might not be appropriate to the study of Congressional polarization.

Although most scholars assume the existence and structure of ideology, Downs argued that the creation of party ideologies, as well as their tendency to diverge, derives from imperfect knowledge in the electorate \cite{downs1957b}.  In a complex world, Downs reasoned, voters demand a method to reduce information costs associated with making electoral decisions.  Motivated by the desire to increase vote shares \cite{mayhew1974}, the parties supply ideologies in response to this demand.  Importantly, Downs argued that ideologies reduce information costs for voters primarily by highlighting differences between parties.  He therefore concluded that `parties cannot adopt identical ideologies, because they must create enough product differentiation to make their output distinguishable from that of their rivals, so as to entice voters to the polls' (Downs 1957b:142)\nocite{downs1957b}.  Given incomplete information, it therefore seems that one should expect some level of polarization rather than convergence to the median.  

Additionally, Downs attributed the original position of party ideology to the interests of those present at that party's founding but presumed that successful ideologies acquire power that is independent of any particular interest group (i.e., parties eventually lose control over the ideologies that they create) \cite{downs1957b}. Indeed, parties have {\it imperfect information} about the nature of their own ideology.  Downs also argued that party ideology must remain relatively stable once established in order to retain legitimacy, and that subsequent political actions must offer `persistent correlation' with that ideology in order for it to remain a useful cognitive shortcut \nocite{downs1957b}(Downs 1957b:142).  Moreover, if a party lacks control over its own ideologies, then it becomes more difficult both for it to ensure that its ideology is stable and for it to craft policy that adheres to its ideology.

In this scenario, ideology is a coordination device between the electorate and the parties.  The parties want to provide just enough information to appear stable and trustworthy to voters, and voters want just enough information to make decisions between competing parties.  If voters are unwilling or unable to demand or process more information \cite{popkin1994}, then parties will prefer vague ideologies.  Ambiguity allows competing intra-party factions to appear united to voters who lack the necessary information to expose their contradicting positions, and it gives parties the opportunity to pursue potentially divisive or polarizing policies while maintaining the guise of ideological consistency.  This helps explain, for example, why parties in Congress appear to be radically divided even when the general public seems not to be so \cite{fiorina2008} and why partisans in the electorate continue to support their representatives even when the rest of the public has withdrawn its support \cite{jacobson2005}.

This exposition of Downs \citeyear{downs1957b} calls into question the assumptions that underly many studies of ideology, partisanship, and polarization.  Scholars traditionally assume that individuals have complete, transitive, single-peaked preferences over a small number of independent ideological dimensions, which presumably represent correlative sets of issue positions \cite{poole1997}.  However, given that all actors have imperfect information not only about the composition of these dimensions but also about the location of party and electoral medians along them, one encounters the possibility of two actors holding radically different opinions about the structure of ideology but still thinking that they occupy the same position.  If ideology means different things to different people, then it makes little sense to assume the opposite.  Furthermore, parties have the incentive to provide---and voters have the incentive to consume---only the minimal amount of information necessary to distinguish between candidates for office, and this undermines the assumption that actors have complete and well-ordered preferences over any particular ideological space.

Finally, nearly all rational-choice models assume that actors are able to make better political decisions when given more information \cite{poole2005}.  Indeed, spatial models are rather extreme in this respect: they assume that actors make perfect decisions when they have perfect information \cite{krehbiel1991}.  For politicians, one typically assumes that perfect decisions are those that maximize reelection chances \cite{mayhew1974}.  For voters, the story is different.  Though an incomplete-information setting encourages voters to rely on partisan, ideological, or other cues \cite{popkin1994}, voters in a complete-information setting are capable in theory of associating policies with outcomes (rendering such cues unnecessary).  In a complete-information setting, reelection rates should hinge on the ability of politicians to effectively generate good public policy, ostensibly by processing the maximal amount of policy information \cite{krehbiel1991}.  This assumption is fundamental to dominant theories of committee organization in Congress \cite{shepsle1987}.  However, empirical investigations suggest that committees use information polemically as a way of defending existing partisan positions while appealing to foundational assumptions among voters about the utility of rational and scientific methods \cite{shulock1998}.  In this case, even analyses of scientific policy serve as low-information cues about the validity of what are essentially ideological or partisan positions.

These critiques imply a world in which ideology is a necessary---but frustratingly imprecise---tool of party competition.  Nearly all studies of Congressional polarization treat it as ideological and then measure polarization using traditional assumptions.  Perhaps the most popular of these measures, defined by McCarty, Poole, and Rosenthal (MPR) \citeyear{mccarty2007}, gauges polarization by measuring the Cartesian distance between the mean DW-NOMINATE (which we hereafter call `DW-NOM') scores of political parties \cite{poole1997}.  Calculating these polarization scores requires restrictive assumptions about the nature of ideology in Congres.  DW-NOMINATE assumes the existence of a low-dimensional space with consistent  ideological dimensions over time.  This assumption is made in in order to estimate dynamic ideology scores.  W-NOMINATE relaxes this restriction while maintaining the spatial modeling assumptions \cite{poole2005}.  In both cases, measuring the distance between political party means, furthermore, requires the researcher to assume a particular party system structure.  We compare the modularity measure to polarization scores calculated using both DW- and W-NOMINATE (see section \ref{sec4} for more discussion).

While we acknowledge the utility of NOMINATE-based measures for fitting individual roll-call decisions, we question the value of aggregating ideal-point estimates into measures of system-wide polarization.  In these situations, we reason, it is prudent to also employ measures that hew more closely to observed behaviors without imposing assumptions about their rationality or spatial structure.  We argue that the polarization debate should be more concerned with the identification of relevant political groups and the evaluation of the divisions between them.  By moving to a network framework, we can use the tools of community detection and the diagnostic known as \emph{modularity} to perform both of these tasks using more plausible assumptions.


\section{The Modularity Measure}\label{sec2}

When studying a network, it is often useful or convenient for analysis to partition it into groups.  Network scientists have recently developed a measure called \emph{modularity} \cite{newman2004,newman2006b} that uses information about the ties between each pair of individuals in a network to compare the total strength of ties lying within each group to the total tie strength between individuals from different groups.  Previous work has used modularity to study cohesive groups (typically called \emph{communities}) in legislation cosponsorships networks in Congress (Zhang et al. 2008)\nocite{zhang2008}, committee membership networks in the House of Representatives (Porter et al. 2005, Porter et al. 2007)\nocite{porter2005,porter2007}, and a large variety of other real-world and computer-generated networks (Porter et al. 2009, Fortunato 2010)\nocite{comnotices,fortunato2010}.  Other applications of network analysis have also flowered in the political science literature (see, e.g.,\nocite{huckfeldt1987,fowler2006,fowler2006b,mcclurg2006,balda,koger2009,park2009,ward2011,lazer2011} Huckfeldt 1987, Fowler 2006a, Fowler 2006b, McClurg 2006, Baldassarri \& Bearman 2007, Koger et al. 2009, Park et al. 2009, Lazer 2011, Ward et al. 2011).

`Modular' networks contain groups that have many ties within them but few between them.  Network scientists call such groups `communities' because they form strongly connected subnetworks that, in the extreme, can be nearly separate from other parts of the network (Porter et al. 2009, Fortunato 2010)\nocite{comnotices,fortunato2010}.  Networks with stronger ties within groups and weaker ties between groups are thereby more modular.  Conceptually, this is exactly what one means when claiming that groups are polarized.  This operationalization of polarization in roll-call votes allows us to quantify the number of cohesive groups (i.e., communities) in a legislature, quantify the strength of division between such blocs, identify which individuals are likely to belong to each cohesive group, and quantify the position of individuals within their groups.  

We employ multiple community-detection algorithms to identify groups that maximize modularity for each roll-call network for both the Senate and the House of Representatives in the 1st--109th Congresses.\footnote{We briefly discuss these procedures in section \ref{detect} of the online supplementary information.}  Using regression analyses, we find that maximum modularity is a significant predictor of future majority party changes in the House (and approaches significance in the Senate).  Additionally, we find several periods in American history---most notably, during the 75th--95th Congresses from 1937 to 1979---in which a large discrepancy exists between formal party divisions and real voting coalitions.  We hypothesize that such discrepancies, and the corresponding changes in maximum modularity, might serve as an early warning signal for changes in the partisan composition of Congress (perhaps due to a failure of parties to coordinate with voters on the definition of ideology).  As a preliminary test of this hypothesis, we use modularity values in Congress $t$ to predict changes in the majority party for Congress $t+1$.  We find a non-monotonic relationship between modularity and the stability of the majority party in both chambers of Congress.  At low levels of modularity, there appears to be little impetus to coordinate a change in majority control; at high levels, there is little the minority can do to overcome the majority's cohesion.  In both of these cases, majority-party switches are infrequent.  However, at medium levels of modularity, there is a mix of impetus and relaxed majority cohesion, yielding a party system that is significantly less stable.  We call this interpretation the `partial polarization' hypothesis.

Importantly, our analysis helps us begin to explain why partially-polarized Congresses exhibit the greatest instability.  Using individual-level diagnostics associated with modularity, we identify the legislators who are most polarizing (via a quantity that we call `divisiveness') and those who align most closely with their group (via a `solidarity' diagnostic).  We show that divisiveness has a negative impact on individual reelection chances but that the effect is mitigated for polarizing legislators who exhibit strong solidarity with their group.  This, in turn, yields instability in partially-polarized Congresses.  Our analysis corroborates previous findings that legislators must balance partisan and constituency interests in order to remain in office (Canes-Wrone et al. 2002)\nocite{canes2002}, while also providing insight into legislators' group affiliations that are not reflected in formal party labels.


\subsection{Modularity Defined} \label{moddef}

We begin with a common assumption about the nature of roll-call votes: Congressmen who vote with one another are more similar than those whose votes conflict.  We further assume that two Congressmen are more similar when they agree on more roll-call decisions.  These assumptions underly all investigations of roll-call voting blocs (Anderson et al. 1966)\nocite{aww1966} from Rice's \citeyear{rice1927} identification of blocs in small political bodies to Truman's \citeyear{truman1959} case study of the 81st Congress, early investigations of policy dimensions by MacRae \citeyear{macrae1958} and Clausen \citeyear{clausen1973}, and more quantitative analyses by Poole and Rosenthal \citeyear{poole1997} and others (Clinton et al. 2004)\nocite{clinton2004}.  

The method of identifying communities that we employ is philosophically similar to the cluster analyses employed by Rice and Truman, but those authors had limited computing power at their disposal and lacked an objective method for evaluating the quality of the communities that they obtain.  However, because of a wealth of conceptual and algorithmic advances from the past decades \nocite{comnotices,fortunato2010}(and especially from the past 10 years; Porter et al. 2009, Fortunato 2010), we have not only an appropriate measure (modularity) but also good computational algorithms to partition networks into communities in order to maximize it.  In the remainder of this section, we define modularity and describe the methodology that we use to generate modularity scores.\footnote{We provide more details of this process in section \ref{detect} of the online supplementary information.}  

Using roll-call data compiled by Poole and Rosenthal \citeyear{poole1997,voteview}, we generate a network in the form of an adjacency matrix \cite{wasserman1994} that describes voting similarities among legislators in a single Congress of the House of Representatives or Senate.\footnote{Defined in footnote 1 and discussed in detail in section \ref{adjmat} of the online supplementary information.}  This is done in similar fashion to the assembly of agreement score matrices in Poole \citeyear{poole2005}.  We study the 1st--109th Senates and Houses, so we consider 218 networks in total.  We represent each of these networks by an $n \times n$ matrix ${\bf A}$, where $n$ equals the number of legislators in the body and each element $A_{ij}$ gives the proportion of votes on which two legislators agreed.  The value of $A_{ij}$ represents the weighted strength of connection between legislators.  Having generated the adjacency matrices, we can calculate modularity values for any given partition of these roll-call networks into specified, non-overlapping communities (Porter et al. 2009)\nocite{comnotices}.  

Modularity relies on the intuitive notion that communities in networks should consist of nodes with more intra-community than extra-community ties (Newman \& Girvan 2004, Porter et al. 2009)\nocite{newman2004,comnotices}.  This mirrors our conceptualization of polarization described in section \ref{rr_bg}.  For a given partition of the network into communities, the modularity $Q$ represents the fraction of total tie strength contained within the specified communities minus the expected total strength of such ties. The expected strength depends on an assumed null model.  Here we use the standard Newman-Girvan null model that posits a hypothetical network with the same expected degree distribution as the observed network \cite{newman2006b,newman2006}.  This standard null model implies that modularity is given by the formula
\begin{equation}
	Q = \frac{1}{2m}\sum_{ij}\left[A_{ij}-\frac{k_{i}k_{j}}{2m}\right]\delta(g_i,g_j) \equiv \frac{1}{2m}\sum_{ij} B_{ij} \delta(g_i,g_j)\,, \label{one}
\end{equation}
where $m=\frac{1}{2}\sum_i k_i$ is the total strength of ties in the network, $k_i=\sum_j A_{ij}$ is the weighted degree (i.e., the strength) of the $i$th node, $g_i$ is the community to which $i$ is assigned (and similarly for $g_j$), and $\delta(g_i,g_j) = 1$ if $i$ and $j$ belong to the same community and $0$ if they do not.  In equation (\ref{one}) we have defined a modularity matrix ${\bf B}$ with components $B_{ij} = A_{ij}-\frac{k_{i}k_{j}}{2m}$.

Modularity evaluates the quality of community partitions, implying that partitions with higher modularity are, by our conceptualization, more polarized.  However, it remains for us to determine the community partition that maximizes modularity for each Congress.   We call this the  `maximum-modularity partition', though strictly speaking no partition can ever be proven to be the global optimum without computationally-prohibitive exhaustive enumeration \nocite{np}(modularity maximization is an NP-hard problem; Brandes et al. 2008).  As discussed in the online supplementary information, we consider a variety of computational heuristics in our optimization of modularity.


\subsection{Modularity at the Individual Level} \label{ccdef}

We also consider individual-level diagnostics associated with modularity: \emph{divisiveness}  can be used to identify the extent to which individual legislators potentially contribute to system-wide modularity (polarization), and \emph{solidarity} can be used to measure the alignment of individual legislators to their communities.  Calculating divisiveness and solidarity allow us to explore hypotheses about the relationship between the behavior of individual legislators and outcomes of interest (such as reelection rates).

Mathematically, the \emph{divisiveness} $|{\bf x}_i|$ of legislator $i$ is obtained from \cite{newman2006b}
\begin{equation}
	|{\bf x}_i|^2 = \sum_{j=1}^p (\sqrt{\lambda_{j}}U_{ij})^2\,, \label{two}
\end{equation}
where $p$ is the number of positive eigenvalues $\lambda_j$ of the modularity matrix ${\bf B}$ and the matrix element ${\bf U}_{ij}$ is the $i$th component of the $j$th (normalized) eigenvector.
That is, ${\bf x}_i$ is a p-dimensional vector (with $j$th element equal to $\sqrt{\lambda_j}U_{ij}$) and the magnitude $|{\bf x}_i| $ of this node vector measures the potential positive impact on aggregate modularity from legislator $i$.\footnote{The divisiveness $|{\bf x}_i|$ is known as `community centrality' in the networks literature \cite{newman2006b}.}

The divisiveness measure uses the roll-call adjacency matrices to estimate the potential effect that each individual legislator has on the aggregate polarization of his/her legislature, but it need not say anything about the alignment of that legislator's voting behavior with that of his or her own group.  Estimating alignment requires us to compare the divisiveness measure with the associated community vector  ${\bf X}_k = \sum_{i \in c^k} {\bf x}_i$\,, where we have summed over all node vectors corresponding to legislators assigned to the $k$th community $c^k$. We can then calculate the \emph{solidarity}\footnote{From a networks perspective, one might wish to use the name `community alignment' for the solidarity.} $\cos \theta_{ik}$, where $\theta_{ik}$ is the angle between the node vector ${\bf x}_i$ and the community vector ${\bf X}_k$.  When the solidarity is close to $1$, the legislator and community are in strong alignment; when the solidarity is close to $0$, however, the legislator is not strongly aligned with his or her community \cite{newman2006b}.

p
\subsection{Modularity and NOMINATE-based Measures} \label{sec3}

Let us first consider the number of communities revealed by the modularity procedure.  Theoretical models suggest that single-member districts should yield a two-party system \cite{duverger1954,cox1997}, so we expect most Congresses to achieve their maximum modularity when partitioned into two communities.  However, we find three or more communities in 35 of 109 Houses and in 67 of 109 Senates.  We tended to obtain more communities when maximum modularity is low.  Maximum modularities in Congresses partitioned into three or more communities are on average $0.045$ lower in the House and $0.066$ lower in the Senate than maximum modularities in two-community Congresses.  These differences are both significant ($p < 0.0001$) in one-tailed $t$-tests.  When we constrain our focus to those Houses and Senates in which the third-largest community is larger than the size difference between the two largest communities, we find three or more communities in 11 Houses and 31 Senates.  We provide descriptive statistics for these Congresses in section \ref{sup3com} of the online supplement and a historical example from the 19th century in section \ref{19th}.

\begin{figure}[htbp]
\centerline{\it{Panel A: House}}
\centerline{
\includegraphics[width=6.75in]{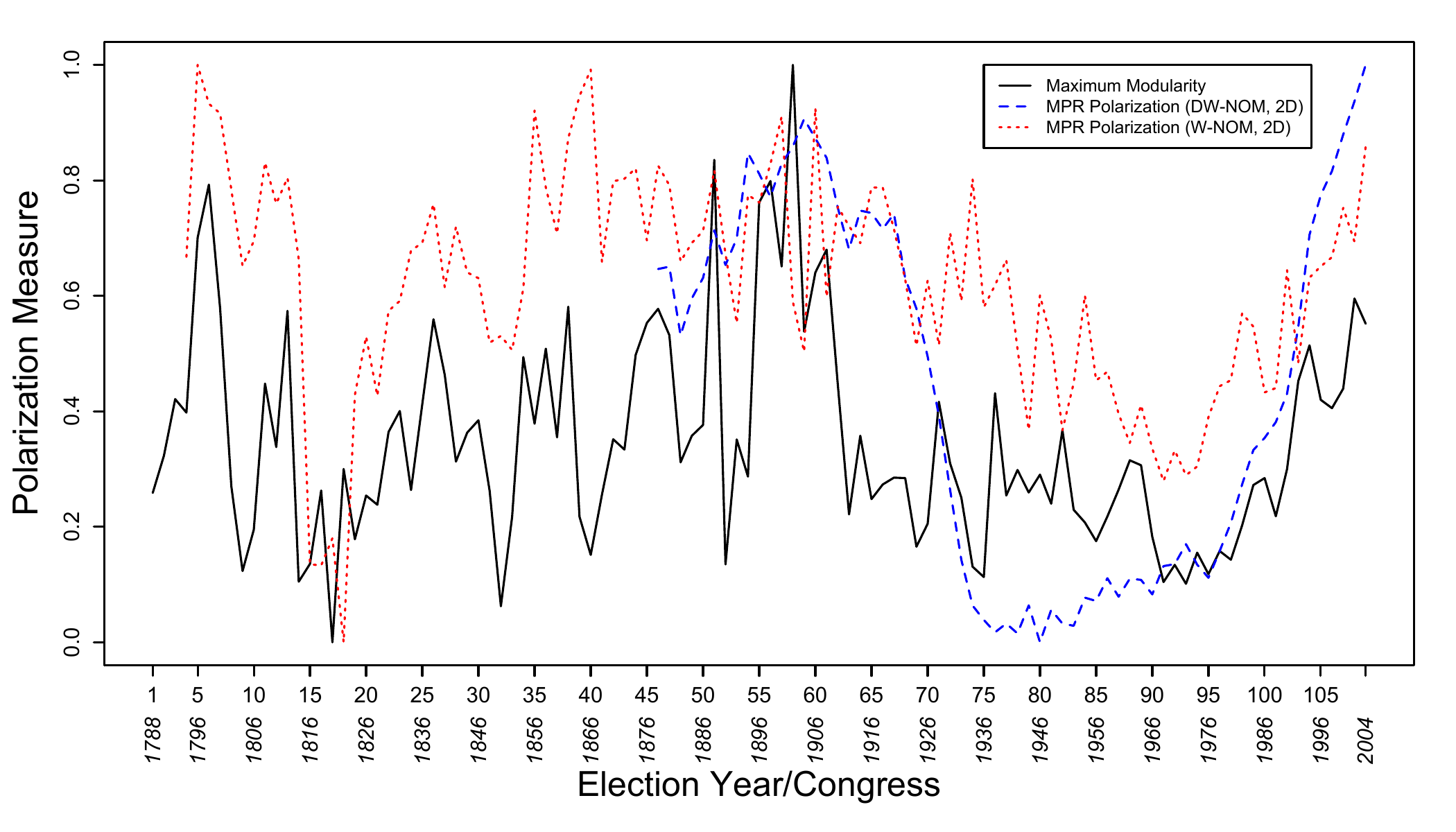}
}

\centerline{\it{Panel B: Senate}}
\centerline{
\includegraphics[width=6.75in]{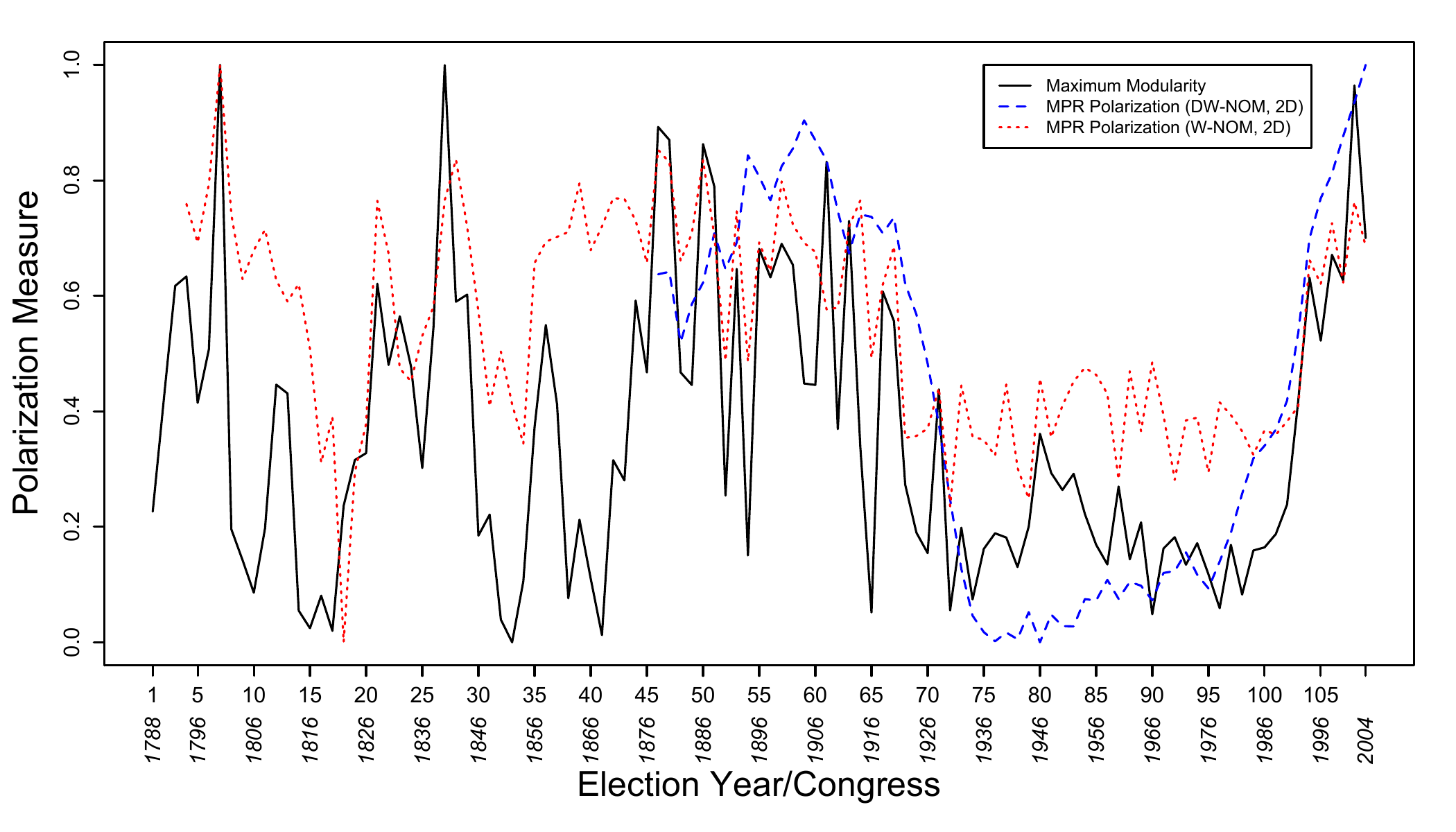}
}
\caption{[Color online] Longitudinal comparison of modularity and MPR measures in the House (Panel A) and Senate (Panel B). Each measure has been rescaled to $[0,1]$ for visual convenience.  Maximum modularity lies in the interval $[0.039,0.364]$ in the House and $[0.061,0.285]$ in the Senate.}
\label{conglong}
\end{figure}

In figure \ref{conglong}, we compare maximum-modularity values to the NOMINATE-based measure used by McCarty, Poole, and Rosenthal \citeyear{mccarty2007} and described in section \ref{rr_bg}.  Following the advice of Aldrich et al. \citeyear{aldrich2004}, we calculate this measure using two dimensions of DW-NOM.  For reference, we include a calculation of the MPR measure using two-dimensions of W-NOMINATE (which we hereafter call `W-NOM') as well.  For comparability, we rescale both modularity and the NOMINATE-based measures to lie in the interval $[0,1]$.

\begin{center}
$<$FIGURE \ref{conglong} ABOUT HERE$>$
\end{center}

The modularity values in figure \ref{conglong} are consistent with several stylized facts about polarization.  Most notably, they capture the spike in polarization associated with the end of Reconstruction as well as the well-documented modern spike (McCarty et al. 2007)\nocite{mccarty2007}.  They also show a lull, corresponding to the era of party decline during the 75th--95th Congresses (1937--1979) \cite{coleman1996b}.   Interestingly, the DW-NOM version of the MPR measure suggests a much lower level of polarization over this period than does modularity or the W-NOM version.  Further, the DW-NOM-based measure derives much of its visual impact from its limitation to post-Reconstruction Congresses.  The modularity and W-NOM-based measures show that modern-day polarization is high but not to a greater extent than what seems to be the case in many other periods.  The low-modularity period of the 75th--95th Congresses appears to be the exception rather than the rule.  

Another difference between modularity and the MPR measure is the year-to-year variation.  DW-NOM assumes that legislators always remain in the same voting bloc and allows their ideology to move in only one direction over time, resulting in a time series that is smoother than that for the modularity or W-NOM measures.\footnote{The W-NOM version also suggests higher levels of polarization than maximum modularity for most Congresses.  We have no theory to explain this finding but stress that the W-NOM measure is not a significant predictor of majority-party switches (see table \ref{modlongtab})}  As we discuss below, we believe that allowing such year-to-year variation is informative.

\begin{figure}[htbp]
\centerline{
\includegraphics[width=6.75in]{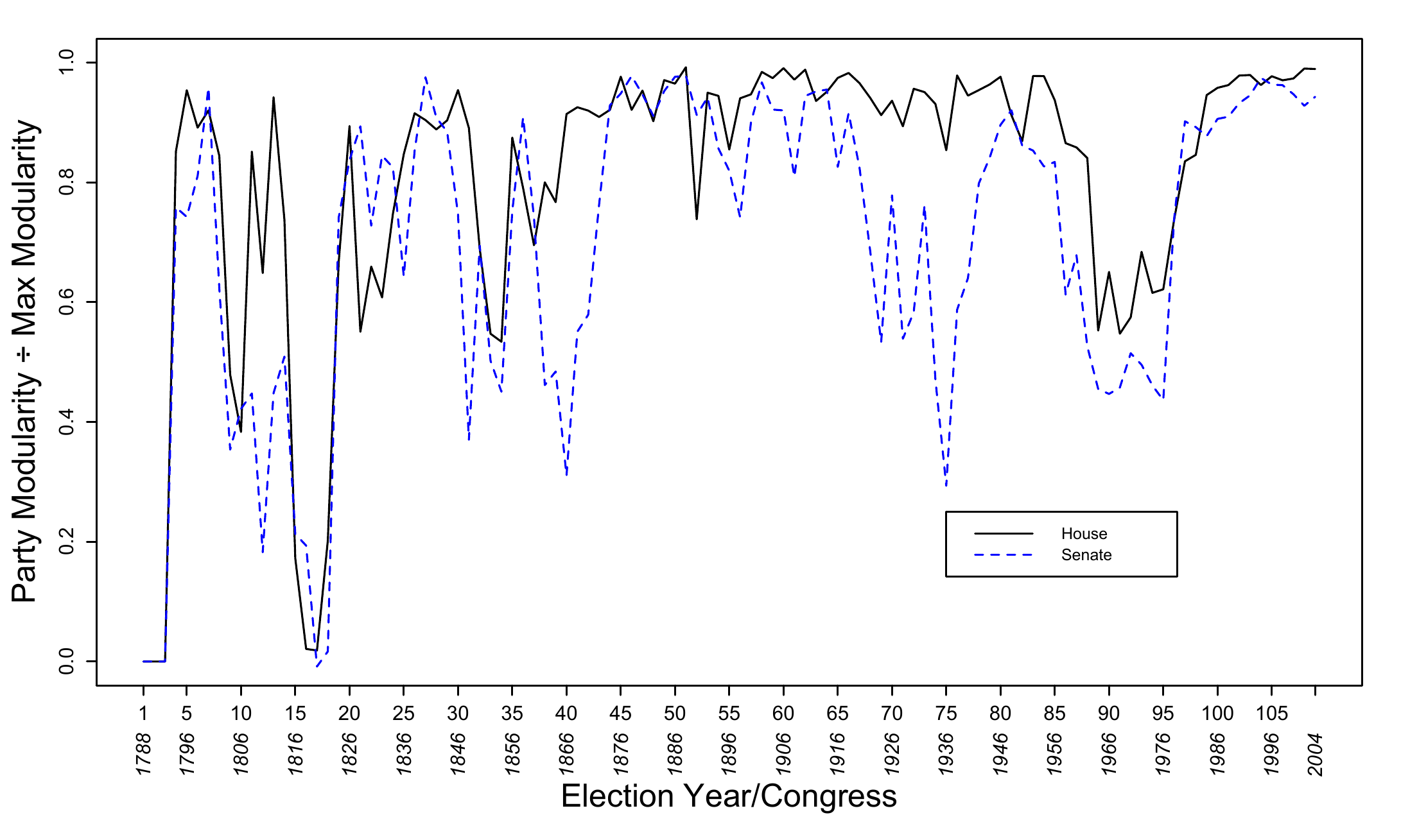}
}
\caption{[Color online] Longitudinal plot of party modularity divided by maximum modularity for both the House and Senate.  The contribution of party to maximum modularity varies considerably over time, particularly in the Senate, suggesting that polarization in Congress is usually, but not always, driven by formal party divisions.}
\label{partyfrac}
\end{figure}

In addition to maximum modularity $Q$, we calculate party modularity $P$, which is the modularity obtained from the network partitioned so that legislators are assigned to groups that contain only members of the same party.  In figure \ref{partyfrac}, we report $P/Q$, which represents the relative contribution of formal party divisions to total polarization.  In periods in which polarization is predominantly partisan, one finds that $P \approx Q$.  When $P$ is substantially less than $Q$, however, community divisions other than party better explain polarization.  The party partition captures the vast majority of the maximum modularity in all modern Houses, with the 85th--95th Congress period (1957--1979) serving as a notable exception.  Party importance varies more in the Senate, where it oscillates from one Congress to the next between the 67th (1921) and 75th (1937) Congresses and is again a smaller fraction of modularity during the 85th--95th Congresses.

\begin{center}
$<$FIGURE \ref{partyfrac} ABOUT HERE$>$
\end{center}

Although the DW-NOM MPR measure suggests a substantially lower polarization level for the 75th--95th Congresses than does the modularity measure, party modularity nearly equals maximum modularity for the 75th--84th Houses before dropping off during the 85th--95th, suggesting that the importance of party did not start to wane until the 85th Congress (some 20 years after the decline in the DW-NOM measure).  Our results suggest that extra-partisan coalition tensions generated heightened polarization during this period that is not captured by DW-NOM, even in two dimensions.\footnote{Regression results, which we present in section \ref{sec4}, provide some support for this finding in the House but not in the Senate.}

The existence of a disparity between party allegiance and voting behavior is unsurprising. Many studies show that parties have reorganized throughout history (Merrill et al. 2008)\nocite{merrill2008}.  These realignments represent changes in the formal allegiances of members of Congress. It is reasonable to assume that such changes are costly to politicians \cite{downs1957b,cox2005}, so they are unlikely to be undertaken without substantial prior effort to salvage the existing party order.  As party bonds disintegrate, we reason that some legislators seek to preserve alliances while other (opportunistic) legislators seek new alliances that reflect (or perhaps help create) a different order \cite{riker1986}.  

The existing measures of polarization based on DW-NOM are ill-equipped to identify these shifts for 
three reasons.  First, they assume a party-system structure to orient their legislators in space, and this assumption might mask the importance of intra-party communities.  Second, DW-NOM is weighted dynamically, which constrains the spatial movement of legislators over time to a single direction.  This restriction allows one to identify ideal points on a consistent spatial metric over time \cite{poole2005}.  Cox \& Poole defend this constraint by noting many legislators have changed parties over their careers but that none have changed back \cite{cox2002}.  This defense is justifiable if exogenously-defined groups, such as parties, sufficiently capture group dynamics \cite{poole1997}, but our results suggest otherwise.  Third, these measures rely on strict assumptions about the nature of ideology \cite{downs1957b} that might be more appropriate for fitting individual roll-call votes than for examining the effect of group dynamics on formal party divisions.


\subsection{Changes in Group Dynamics} \label{group}

One clear indicator of a formal power shift in Congress is a change in the majority party.  When a new majority party is elected, one can normally point to major policy failure on the part of the previous majority.  One important way for a majority party to remain effective---and for its brand to remain strong---is for its caucus members to coordinate on a policy agenda.  The House and Senate majorities resolve their coordination problems through various institutional means, such as the delegation of agenda control to leaders and the appointment of party whips \cite{rohde1991,cox2005}.  Willingness to coordinate depends on a variety of forces, including electoral pressure, ideological cohesion, and career ambition \cite{aldrich2001}.

When party membership poses electoral risk, members hedge their bets by seeking extra-partisan coalitions. Modularity captures this dynamic, showing the emergence of third (non-party) communities and the evolution of party-dominated communities into more heterogenous groups.  The less that communities in Congress reflect party labels, the more likely that interest groups, party organizations, and ultimately voters notice the gap.  If these political actors support the party, then the legislator might be replaced and the party system might thereby be preserved.  However, if they support the legislator, then he/she might either switch parties or attempt to bring his/her party closer to his/her district's preferences.  When party positions shift, we reason, it becomes more difficult for parties and voters to coordinate on ideology.  Thus, electoral volatility increases and changes in formal groups, such as majority party switches, become more likely.

In order to test the ability of maximum modularity to predict changes in the majority-party in Congress, we examine the values of modularity in section \ref{sec4}.  We then conduct individual-level analyses in section \ref{comcent} to examine the ability of divisiveness and solidarity (which we defined in section \ref{ccdef}) to predict the electoral fates of House members.  In both cases, we compare regressions using the modularity measures to similar specifications that use NOMINATE-based measures as their independent variables.


\section{Majority-Party Switches} \label{sec4}

\begin{figure}[htbp]
\centerline{\it{Panel A: House}}
\centerline{
\includegraphics[height=3.75in]{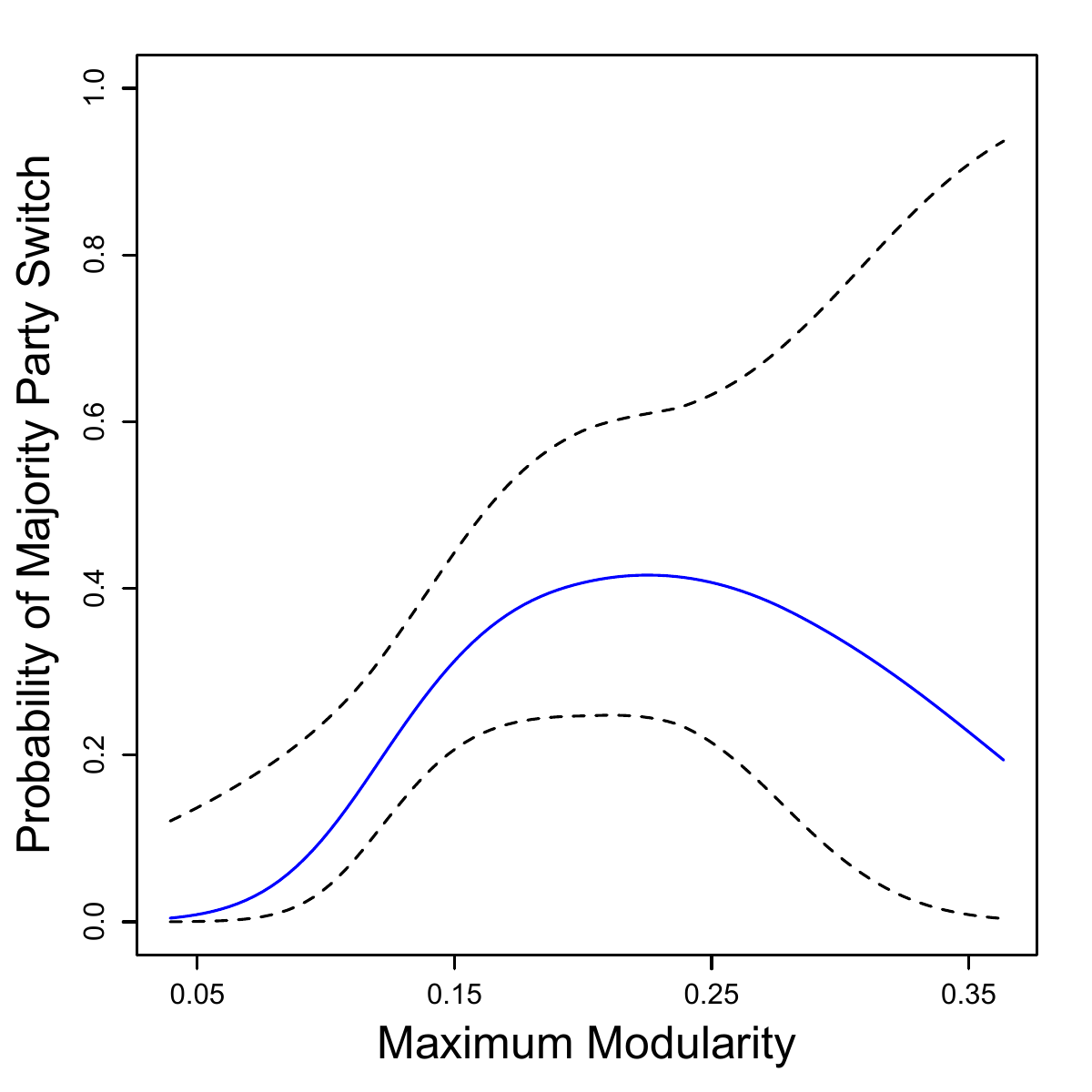}
}
\centerline{\it{Panel B: Senate}}
\centerline{
\includegraphics[height=3.75in]{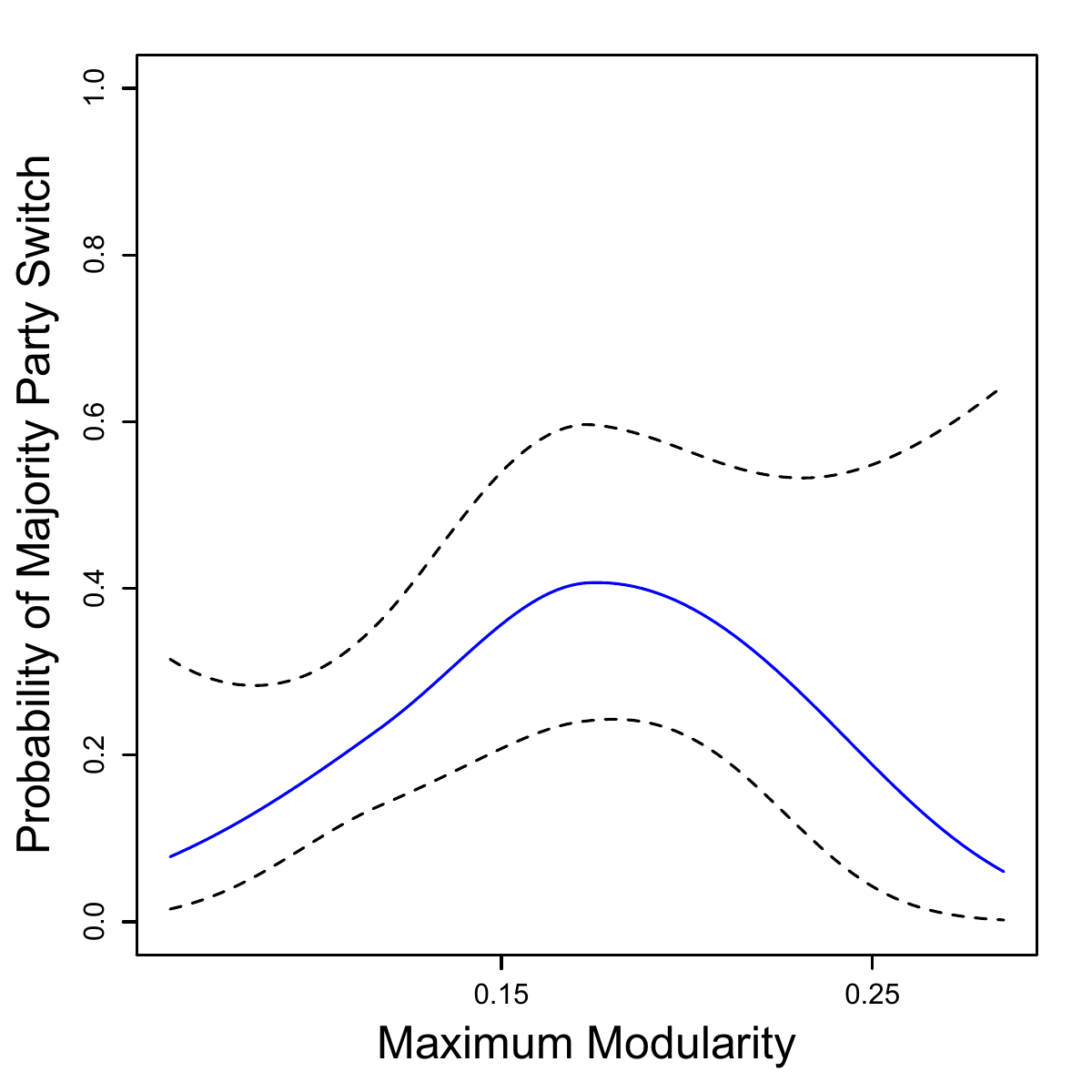}
}
\caption{[Color online] LOESS plot of maximum modularity in Congress $t$ versus majority party change in Congress $t+1$ for the House and Senate.  Majority party changes are most probable during medium-modularity Congresses.}
\label{congloess}
\end{figure}

We begin to explore the relationship between maximum modularity in Congress $t$ and a majority party switch in the next Congress ($t+1$) using locally weighted polynomial regression (LOESS) \cite{loader1999}.  Our analyses demonstrate that changes in majority-party switches are most common when polarization is moderate, and they are relatively uncommon when polarization is low or high.\footnote{The medium-modularity range is approximately $[0.15,0.30]$ for the House, and $[0.15,0.23]$ for the Senate (see figure \ref{congloess}).  We show associated plots in section \ref{supcloess} of the online supplementary information.}  This non-monotonic relationship suggests that it is appropriate to include both linear and squared modularity scores in multivariate regressions.


\subsection{Data, Analyses, and Results} \label{data}

We compiled a time-series data set that covers each of the 4th--109th Congresses.\footnote{The accompanying economic data that we gathered were not available for the first three Congressional sessions.}  The data set contains both the key independent variable (\emph{maximum modularity}) and the key dependent variable (\emph{majority-party switches}) for each House and Senate.  `Majority-party switches' is a binary variable that takes the value $1$ if a switch occurred as a result of the previous election and a $0$ if it did not.  Using information provided in Kernell et al.~\citeyear{kernell2009}, we identified 27 switches in the House and 26 switches in the Senate.\footnote{We include a table of these switches in section \ref{switches} of the online supplementary information.}

We control for economic indicators such as gross domestic product (GDP), consumer price index (CPI), and national debt (as a percentage of GDP) (Historical Statistics of the United States, 2009)\nocite{hsus2009}.  We also include indicator variables for divided government, midterm Congress, and Republican or Democratic majorities.  We included the first two variables to control for the impacts of presidential races on Congressional electoral outcomes and the last two to capture any peculiar effects of a stable two-party system.  Finally, we include an indicator variable for Congresses with three or more communities, expecting that these Congresses are likely to be particularly unstable.

We conducted logistic regression analyses using several specifications.  We first examined the relationship between future majority-party switches and maximum modularity and its squared term, while controlling for the lagged dependent variable and the presence of three (or more) communities.  We subsequently added structural and economic control variables for Congress $t$ and retested the model for majority-party switches in the next Congress ($t+1$).  To aid comparison with NOMINATE-based measures, we report similar specifications using both W- and DW-NOM-based MPR measures of polarization.   MPR analyses are necessarily limited to Congresses 46-109 (the time period over which DW-NOM is calculated), so we conducted modularity regressions over this time period as well.  We present our results in table \ref{modlongtab}.

\begin{center}
$<$TABLE \ref{modlongtab} ABOUT HERE$>$
\end{center}

The regression results show a clear relationship between modularity and majority-party shifts in the House.  In all four specifications, modularity is significant and positive ($p < 0.05$), and its squared term is significant and negative ($p < 0.05$). In the Senate, results are weak.  Modularity approaches significance ($p < 0.1$) in only two of the four models, and the squared term approaches significance in only one.  Neither the DW- nor W-NOM MPR measures are significant in any specification in either chamber.

The existence of a non-monotonic relationship between modularity and House majority switches has important implications for the study of legislative organization and party dynamics.  With some caution, we offer some preliminary explanations for these results in the following section.  We refer to these results and explanations as our `partial polarization' hypothesis.


\subsection{Discussion} \label{discuss}

We believe that the instability of partially-polarized Houses might be driven by the strategic behavior of legislators, candidates, and other partisans as they attempt to coordinate with low-information voters.  These dynamics are most easily explained by dividing Congresses into three categories: those with low, medium, and high modularities.  We observe that low- and high-modularity Congresses tend to have stable majorities, whereas the majorities in medium-modularity Congresses are less stable.

In low-modularity Congresses, communities tend to be weak and are presumably less informative to political elites and the general electorate.  Recall from section \ref{sec3} that Congresses with more than two communities tend to have lower modularity than those with two communities.  If one imagines Congressional activity as a coordination problem, then low-modularity Congresses are those in which coordination takes place between different coalitions on different issues and in which mechanisms to aggregate preferences within groups have little power.  In such an environment, coordination costs are likely to be high, and individual legislators might see little benefit in group alliances \cite{olson1965}, which could result in committee rule \cite{shepsle1987} or gridlock \cite{binder1999}.  Electoral institutions \cite{herrnson2004} also give Congressmen the incentive to pursue particular benefits for their districts in order to win reelection (regardless of collective impact).   In such an environment, coordination likely occurs through logrolling and the exchange of district-level benefits.

High-modularity Congresses have the opposite problem: communities are well-defined and usually party-oriented.  Presumably, legislators have solved their coordination problem by coalescing into voting blocs that reduce the costs of governing and improve the value of ideological signals to the electorate.  Such efficiency comes with a corresponding loss in voting freedom, as the electoral costs of defecting from a community increase.  Donors, lobbyists, activists, and elites who have invested in the existing community structure might be less willing to support defectors, which impairs a Congressman's ability to fundraise and decreases his/her chance of reelection.  Defection might also muddle the ideological signal to voters, which could in turn decrease turnout or encourage the consideration of challengers.  Consequently, pressure to conform might explain the dearth of majority switches in these Houses.

Medium-modularity Congresses reveal environments that are subject to potential flux.  Such Congresses might represent a highly modular environment that is in the process of breaking down or a poorly-structured environment in the process of consolidating.  When group structures exist but are not well-established, politicians have a strategic incentive to develop and control stable communities that will convey more effective signals to voters.  The strategic behavior of legislators, in turn, causes communities to fracture and reassemble; meanwhile, voters attempt to make sense of the more complex environment.

We investigate the strategic incentives of individual legislators in the following section by using divisiveness and solidarity.  These concepts connect modularity and group dynamics directly to the electoral fortunes of individual legislators.


\section{Reelections in the House} \label{comcent}

We begin with insights that arise from our analyses in section \ref{discuss} on the effect of the maximum-modularity value on changes in the majority party in Congress.  Our `partial polarization' hypothesis suggests that medium levels of maximum modularity might lead to instability in Congressional blocs, with some alliances breaking down and others being forged.  At the individual-level, this instability should be reflected in the electoral successes and failures of legislators.  Here we explore the relationship between maximum-modularity, its associated individual-level quantities (namely, divisiveness and solidarity), and reelection outcomes in the House. We start this exploration by conducting a series of two-dimensional LOESS regressions.\footnote{We show associated plots in section \ref{supdivsol} of the online supplementary information.}

In our first regression, we find evidence for an interactive effect between modularity and divisiveness as they impact reelection.  Divisive legislators in medium-modularity Houses have the highest rates of electoral failure but are more successful when modularity is low or high.  In low-modularity Houses, we suspect that this arises because group solidarity is a less valuable cue for voters when group structures are weak.  In high-modularity Houses, we suspect that divisiveness is only successful in combination with strong group solidarity, as groups are highly salient in these Congresses and members are likely to be penalized for defection.  In medium-modularity Congresses, the legislative environment appears to be more complex, as both Congressmen and voters have poorer information about the structure and salience of communities.  This results in coordination failures between Congressmen, parties, and voters \cite{downs1957b}, which in turn leads to lower reelection rates.

We conduct a second regression to illuminate the impact of solidarity and divisiveness on reelection.\footnote{To aid interpretation, we report correlations between divisiveness and solidarity in section \ref{supdivsol} of the online supplementary information.  We also report examples of the various legislator types.}  We find that highly divisive Congressmen suffer in their electoral prospects unless they also have high group solidarity, providing further tentative support for our `partial polarization' hypothesis.  Significant numbers of legislators with both high divisiveness and high solidarity in a Congress are associated with high modularities (that is, they have a high potential contribution to modularity and that contribution is realized by their strong alignment with identified groups; see also the additional regressions in section \ref{supdivsol} of the online supplement).  In the absence of such strong solidarity, the aggregate maximum modularity may only reach the medium-modularity level; legislators in such Congresses possibly use votes to form coalitions (yielding high divisiveness) but are not always successful in forming cohesive groups in line with the full body of their votes (i.e., low solidarity).  Electoral failure in this case could stem from the loss of activist support \cite{aldrich1995} or a coordination failure with voters \cite{downs1957b}.

From these regressions, we derive three testable hypotheses that lend support to our broader `partial polarization' hypothesis.  First,  increasing divisiveness causes a \emph{decrease} in reelection probability.  Second, increasing solidarity \emph{increases} reelection chances.  Finally, the impaired electoral chances of highly-divisive legislators can be mitigated if their divisive behavior is consistent with the voting behavior of their community.  In other words, we expect a positive association between electoral success and an \emph{interaction} between divisiveness and solidarity (divisiveness $\times$ solidarity).  We test these three hypotheses in the next subsection.


\subsection{Data, Analyses, and Results}

In this section, we test our three hypotheses concerning reelection to the House of Representatives. Our dependent variable is reelection to the House (1 for success, and 0 for defeat).  We exclude legislators who do not participate in the general election.  Our explanatory variables are \emph{divisiveness}, \emph{solidarity}, and their interaction.  We have rescaled divisiveness to the interval $[0,1]$ to make the regression results easier to interpret (solidarity lies in this interval by definition).  

We also include several control variables in our specifications.  At the Congress level, we control for presidential election years and divided government using indicator variables.  At the individual level, we control for party (indicators for `Democrat' and `Republican'), preference extremity (absolute value of first-dimension DW-NOM), seniority (number of Congresses served), party unity (as defined in Poole 2005\nocite{poole2005}), previous margin of victory (percentage of votes), and district-level partisanship.\footnote{District-level partisanship is estimated by multiplying the most recent Democratic Presidential vote (percentage) for a district by the party indicator variables.}  We pool data for the 56th---103rd Houses (1899---1995) for our analyses.\footnote{Covariate data were compiled by Keith Poole in collaboration with Andrew Scott Waugh.}

We require a model that can estimate appropriate standard errors for our explanatory variables while accounting for the nature of the data.  To do this, we employ mixed-effects logistic regression model that allows us to account for both Congress-level and individual-level components as random effects by calculating random intercepts for these variables \cite{gelman2007}.  We draw the random intercepts from a Gaussian distribution with mean $0$ and standard deviation equal to that of the variable.  The functional form of the model resembles a traditional logit model with the random effects as additional parameters:
\begin{equation}
	Pr(y_{i} = 1) = \mbox{logit}^{-1} (\alpha^{\mbox{legislator}_{i}} + \alpha^{\mbox{Congress}}_{i} + \beta^{\mbox{fixed}}_{i} + \epsilon_{i})\,. \label{melogit}
\end{equation}

We evaluate four mixed-effects logistic regression specifications and report our results in table \ref{indmelogit}.  In the first specification, we regress divisiveness and solidarity against the reelection indicator.  We find that divisiveness has a significant negative impact on reelection chances and that group solidarity has a significant positive impact.  In the second specification, we also include the interaction of solidarity and divisiveness (Divisiveness $\times$ Solidarity).  This model maintains the finding that divisiveness is associated with decreased reelection probability, and it also suggests that the combination of divisiveness and solidarity has a significant positive impact on reelection.  Although the sign of solidarity flips from positive to negative, the aggregate effect of solidarity must include the interaction term.\footnote{The interaction variable projects the node vector onto the group vector of its associated community, thereby indicating the actual contribution to aggregate modularity that is made by assigning the legislator to that specific group \cite{newman2006b}.}  At even moderately low levels of divisiveness (i.e., rescaled values of 0.2 or lower), the effect of the interaction term exceeds the main solidarity effect, suggesting that most legislators benefit from increased solidarity with their community.   The third and fourth specifications include the aforementioned Congress-level and individual-level controls, and they yield similar results.

\begin{center}
$<$TABLE \ref{indmelogit} ABOUT HERE$>$
\end{center}


\subsection{Discussion}

The negative influence of divisiveness suggests that legislators suffer a penalty for polarizing votes unless they also exhibit high community solidarity, and that this is particularly true in high-modularity environments.  We hypothesize that divisiveness without solidarity not only disconnects legislators from needed elite support but also complicates the decisions of rationally-ignorant voters.  Conversely, legislators cannot be highly divisive while simultaneously maintaining independence from a coalition.  Only by appropriately balancing group solidarity and individual divisiveness do Congressmen maximize their chances at reelection.

Our individual-level results support our Congress-level findings and show significant differences in the value of communities across different levels of polarization.  In partially-polarized (i.e., medium-modularity) Congresses, legislators face complex environments in which their alliance choices are subject both to greater error and to greater risk.  In such environments, they must balance community cohesion with concerns for constituents, activists, and others who impact their electoral fates.


\section{Conclusions} \label{sec5}

Researchers have long sought to separate the effects of party on voting behavior from electoral, interest-group, and other pressures.  Prior studies have assumed the existence and structure of parties \cite{poole1997}---or of alternate mechanisms such as committees \cite{shepsle1987} or institutional veto players \cite{tsebelis2002}---to derive implications for the structure of roll-call voting.  These studies have tended to consider their organizational mechanisms in isolation, which our results suggest might be a mistake.

In this paper, we have used the network-science concept of modularity to provide a novel measure of polarization in Congress.  Using roll-call data, we calculated the community structures that maximize modularity for the 1st--109th Houses of Representatives and Senates.  Such structure includes the membership of each community and also (via modularity) indicates the cohesiveness of communities.  We argue that modularity offers a clearer and more parsimonious measure of polarization than existing measures that are based on spatial-modeling assumptions.  The introduction of modularity and related measures to the analysis of Congressional behavior has the potential to fundamentally alter the study of group dynamics and partisanship in legislatures.

We demonstrate the value of this modularity by demonstrating that there exists a non-monotonic relationship between modularity and majority party switches in the House, which suggests that `partially-polarized' Congresses are more unstable than ones with either low or high levels of polarization.  Similar uses of NOMINATE-based polarization measures fail to replicate this result.  We further investigate the `partial polarization' hypothesis using divisiveness and solidarity, which capture the individual-level impacts of legislative alliances, and we find that these measures are significant predictors of reelections in the House.

Modularity provides a valuable and parsimonious benchmark measure of polarization against which to compare alternate legislative orderings.  By comparing the maximum modularity of a Congress to modularities calculated either using party divisions or using any other exogenously-determined partition, one might be able to identify the conditions under which particular structural arrangements succeed and fail.  This, in turn, might help to disentangle the complex interplay of environmental, ideological, and institutional pressures that impact the structure of Congressional voting.  Our results also suggest that community structure in Congress strongly influences the strategic incentives of political elites to preserve or subvert existing order, and that the value in pursuing a new order depends on the presence of community structures that are neither too strong to break nor too weak to identify.



\section*{Acknowledgements}

We thank Keith Poole for providing roll-call data via {\tt voteview.com} (Poole 2011)\nocite{voteview}. We thank Dan Fenn, Santo Fortunato, A.~J. Friend, Wojciech Gryc, Eric Kelsic, Keith Poole, Amanda Traud, Doug White, and Yan Zhang for useful discussions.  We also thank Kevin Macon, Stephen Reid, Thomas Richardson, and Amanda Traud for use of their code.  ASW, JHF, and MAP gratefully acknowledge a research award (\#220020177) from the James S. McDonnell Foundation.  PJM's participation was funded by the NSF (DMS-0645369).  LP thanks the SURF program at Caltech.


\clearpage{}

\section*{Supplementary Information} \label{supinfo}
\setcounter{section}{7}


\subsection{Generation of Adjacency Matrices} \label{adjmat}

In this section, we describe in detail the process by which we calculated adjacency matrices using House and Senate roll-call data (which we obtained from {\tt voteview.com}) \nocite{voteview}.  For each legislative body, roll calls for a two-year Congress are encoded in an $n \times b$ matrix ${\bf M}$.  Each matrix element $M_{ik}$ equals $1$ if legislator $i$ voted yea on bill $k$, $-1$ if he/she voted nay, and $0$ otherwise.  Because we are interested in characterizing similarities between legislators (rather than direct connections between legislators and bills), we transform the voting matrix into an $n \times n$ \emph{adjacency matrix} ${\bf A}$ whose elements $A_{ij} \in [0,1]$ represent the extent of voting agreement between legislators $i$ and $j$. We define these matrix elements by
\begin{equation}
	A_{ij} =\frac{1}{b_{ij}}\sum_{k} \gamma_{ijk}\,,
\end{equation}
where $\gamma_{ijk}$ equals $1$ if legislators $i$ and $j$ voted the same on bill $k$ and $0$ otherwise, and $b_{ij}$ is the total number of bills on which both legislators voted.  Because perfect similarity between a legislator and him/herself provides no information, we set all diagonal elements to be zero (i.e., $A_{ii} = 0$).  The matrix $A$ thereby encodes a network of weighted ties between legislators, and the weights are determined by the similarity of their roll-call records in a single two-year Congress.  

Following the guidelines of Poole \& Rosenthal \citeyear{poole1997} and Anderson et al. \citeyear{aww1966}, we consider only `non-unanimous' roll-call votes.  A roll-call vote is classified as `non-unanimous' if at least $3\%$ of legislators are in the minority.  For modern Congresses, this implies that a roll-call minority must contain at least $4$ Senators or at least $13$ Representatives to yield a `non-unanimous' vote.  This ensures that our data sets mirror those used by McCarty, Poole, and Rosenthal \citeyear{mccarty2007}, permitting more explicit comparison of our polarization measure with theirs.


\subsection{Community-Detection Heuristics} \label{detect}

In this section, we list and provide brief descriptions of the community-detection heuristics that we used to calculate maximum modularity partitions of the Senate and House networks.  We also indicate references that provide complete specifications of the algorithms.  We note that it is important to consider several such algorithms when studying community structure by optimizing a quality function such as modularity \cite{resolution2007,comnotices,fortunato2010,good2010}.  In the results of our investigation, we partition each network using various community-detection algorithms, and we use in each case the highest-modularity partition that we obtained.

\begin{center}
$<$TABLE \ref{heusums} ABOUT HERE$>$
\end{center}

\begin{center}
$<$TABLE \ref{heusums2} ABOUT HERE$>$
\end{center}

Heuristics 1--5 are all spectral methods of modularity optimization.  Spectral methods use eigenvectors of the modularity matrix ${\bf B}$ that are associated to ${\bf B}$'s largest positive eigenvalues.  Heuristics 1--3 consist of three different implementations in which we use only the leading eigenvector (i.e., the one associated with the largest eigenvalue), so the final partition is obtained from recursive steps involving partitions of some portion of the network into two smaller pieces \cite{newman2006,newman2006b,fortunato2010}.  The difference between the heuristics 1--3 arises from the use of different tie-breaking and fine-tuning procedures, which attempt to improve partitions between the recursive spectral partitioning steps.  In heuristics 4 and 5, we use the two eigenvectors associated with the two largest eigenvalues of ${\bf B}$ \cite{rich2009}:  Heuristic 4 allows only bi-partitioning steps, whereas heuristic 5 allows both bi-partitioning and tri-partitioning steps.  Heuristic 6, known as the `Louvain' method, is a locally greedy modularity-optimization technique \cite{blondel2008,fortunato2010}.  Heuristic 7 employs simulated annealing to maximize modularity \cite{reich2006,fortunato2010}.  

All of the results from heuristics 1--7 have been improved by subsequent application of a Kernighan-Lin algorithm \cite{kerni1970,comnotices,fortunato2010} \citeaffixed{newman2006b,rich2009}{via the specification described in}.  This algorithm takes the community partitions generated by the heuristics and conducts a series of node swaps---moving nodes from one community to another---in order to find higher modularity values.  This is a fine-tuning procedure that can be applied to the partition obtained from any other method.

Heuristic 8 is the walktrap algorithm \cite{pons2005}.  This algorithm starts by partitioning the network into $n$ communities that each contain a single node (i.e., a single legislator).  It calculates a distance between each pair of communities and then begins merging groups by taking random walks between them.  After each merging step, one calculates the modularity score for the current partition.  The algorithm finishes after $n-1$ steps when the nodes have been merged into a single community and reports the highest-modularity partition that it observed during the whole process.

Finally, heuristic 9 is a classical cluster-analysis procedure, which we include for comparison with modularity-maximizing community-detection heuristics.  This `partitioning around medoids' (PAM) cluster-analysis technique is related to the popular $k$-means clustering procedure (which divides a network into precisely $k$ communities), but is more robust: one ordinarily needs to specify $k$ in advance---which is inappropriate for our investigation---but the PAM method allows one to determine an optimum number of communities based on mean silhouette width \cite{kaufman1990}.

In tables \ref{heusums} and \ref{heusums2}, we provide summary statistics about the community-detection heuristics we used.  Note that the modularity values obtained using different modularity-optimization heuristics vary little, especially in the House.  Additionally, neither the cluster-analysis technique nor the walktrap algorithm ever obtain the best result.  Moreover, by using many different computational heuristics, we more confidently sample the complicated modularity landscape to find higher-modularity partitions to employ in our subsequent analysis.  As has been discussed in a recent paper on modularity-optimization in practical contexts (i.e., situations that consider real-world networks) \cite{good2010}, the use of multiple different optimization heuristics is an important protocol to follow.


\subsection{Descriptive Statistics for Congresses with Three or More Communities} \label{sup3com}

In table \ref{3comstats}, we give descriptive statistics for Congresses in which community-detection identifies three or more communities.  In the table, we list all Congresses in which the third-largest community has at least as many legislators as the number in the largest community minus the number in the second-largest community.  For each community in such Congresses, we give the size (i.e., number of legislators), mean divisiveness of legislators, and mean solidarity of legislators.

\begin{center}
$<$TABLE \ref{3comstats} ABOUT HERE$>$
\end{center}


\subsection{A Descriptive Example from the 19th Century} \label{19th}

As we discussed in Section \ref{sec3}, modularity and its associated individual-level quantities can be used to investigate party polarization even during periods of history that include more than two dominant parties.  In this section, we demonstrate the utility of modularity using an illustrative 19th century example.

In the early 19th century, the fledgling party system of the United States was going through a transitional period.  The existing party system, which pitted the dominant Democratic-Republicans against a dying Federalist party, finally broke down in the 18th Congress (1823--1825) as the Democratic-Republicans broke ranks based on their affiliations with national leaders (most notably, John Quincy Adams and Andrew Jackson).  This resulted in a new period, reflected by partisan conflict between supporters of Adams and supporters of Jackson, which lasted until the emergence of the Whigs and the Democratic party in the 25th Congress (1837--1839) \cite{kernell2009}.  The time series of maximum modularity (table \ref{modlongtab}) captures this transition nicely and provides evidence that group structures began to change as early as the 14th Congress---before the Democratic-Republicans divided into the aforementioned camps.  The Adams-Jackson party system finally emerged in the 19th Congress, representing a majority-party switch in both chambers.

One can see using modularity that the breakdown of the Democratic-Republican Party first becomes apparent in the transition from the 13th to 14th Congress (i.e., with the 1814 election).  The second largest negative shift in maximum modularity over the last 200 years occurs during this transition in both chambers ($-0.152$ in the House and $-0.085$ in the Senate).  This decline is particularly interesting given that the country was experiencing a unified Democratic-Republican government and that the Democratic-Republicans held huge majorities in both chambers.  Some of this decline is likely due to the end of the War of 1812 during the 13th Congress.  With the war over, the Democratic-Republicans no longer needed to maintain a united front, which freed legislators to pursue alternate agendas.  

This breakdown yielded a maximum-modularity partition with four communities in the Senate (containing 14, 13, 12, and 5 Senators), and three communities in the House (containing 80, 71 and 44 Representatives), suggesting that Democratic-Republicans in the both chambers were already beginning to explore alternate alliance structures.  In table \ref{3comstats} we see that the mean solidarity scores for the 14th Congress communities are 0.70, 076, 0.66, and 0.82 in the Senate and 0.5, 0.58, and 0.57 in the House.  Compare these to the 13th Congress, in which both chambers have two communities, with sizes of 26 and 20 in the Senate and 117 and 78 in the House, and mean solidarity scores of 0.63 and 0.74 in the Senate, and 0.78 and 0.87 in the House.  While mean solidarity appears not to vary across the 13th-14th Senates, mean solidarity in the House appear substantially lower in the 14th Congress than the 13th.  The decrease in solidarity, coupled with the increase in the number of communities, suggests a weakening of party control over the House in this period.

In the House, this transition becomes further apparent in the 17th Congress (1821--1823), where we again identify three communities when the Democratic-Republican party is nominally whole and maintaining large majorities in both chambers of Congress.  These communities contain 78, 65, and 56 legislators (see table \ref{3comstats}).  By the 18th Congress, the divisions within the Democratic-Republican party become formally acknowledged, as three camps emerge behind the leadership of Adams, Jackson, and William Crawford.  This formal recognition results in a dramatic increase in the maximum modularity of the House compared to the previous Congress, demonstrating the impact that formal party divisions can have on legislator behavior.  

The same basic groups had emerged during the 17th House but had not yet formally consolidated into well-defined camps, as evidenced by their mean solidarity scores (0.39, 0.40, and 0.49).  After this consolidation, however, the cost to coordinate had become lower, with an accompanying increase in maximum modularity and the elimination of the third major community in the modularity-maximizing structure.  In this case, the two largest communities contain 114 and 104 legislators.\footnote{We technically observe 4 communities in the 18th House, but the third and fourth communities contain only 2 and 1 legislators respectively.}  The elimination of the third community despite the emergence of a third party is especially interesting, as it suggests that two of the parties saw the institutional value in coordinating on roll-call votes in order to pursue their own agendas in a majority-rule institution.  This is corroborated by the mean solidarity scores for the two largest communities, 0.62 and 0.61 respectively, which are substantially higher than the three large communities identified in the 17th House.  Thus, the maximum-modularity time series identifies the presence of a coalition government in the 18th House.  By the 19th House, however, the party system reforms into the pro-Adams and pro-Jackson divisions.  A majority-party switch occurs, the pro-Adams party assumes control of both chambers, and a two-community structure emerges.  A significant third community does not emerge again until the 32nd Congress (1851--1853).

We see from this historical example that there is a lag between formal definitions of party and the emergence of coalitions in Congress, and that this lag can work in both directions.  In the 14th Senate and 17th House, we identify four and three communities, respectively, in a period in which only two parties nominally existed.  When a three-party system becomes recognized in the 18th Congress, maximum modularity increases and the number of communities in the House reduces to two.  Modularity thus captures a fascinating distinction between the formal, self-identifying claims of political parties and what coordinating activities are revealed by actual voting behavior.


\subsection{Modularity and Majority Party Switches} \label{switches}

In this section, we present table \ref{partyswitch}, which summarizes the changes in majority party in the United States Congress from 1788-2002.  These switches are used to generate the dependent variable for our Congress-level regressions in section \ref{sec4} of the main text.

\begin{center}
$<$TABLE \ref{partyswitch} ABOUT HERE$>$
\end{center}

\subsubsection{Congress-level LOESS Plots} \label{supcloess}

In this section, we present plots of our Congress-level LOESS regressions \cite{loader1999} to supplement section \ref{sec4} of the text.  Observe the non-monotonic relationship between maximum modularity and majority party switches.

\begin{center}
$<$FIGURE \ref{congloess} ABOUT HERE$>$
\end{center}


\subsection{Divisiveness and Solidarity} \label{supdivsol}

\begin{figure}[ht]
\begin{minipage}[b]{0.5\linewidth}
\centerline{\it{Panel A: Divisiveness and Solidarity}}
\centerline{
	\includegraphics[height=3in]{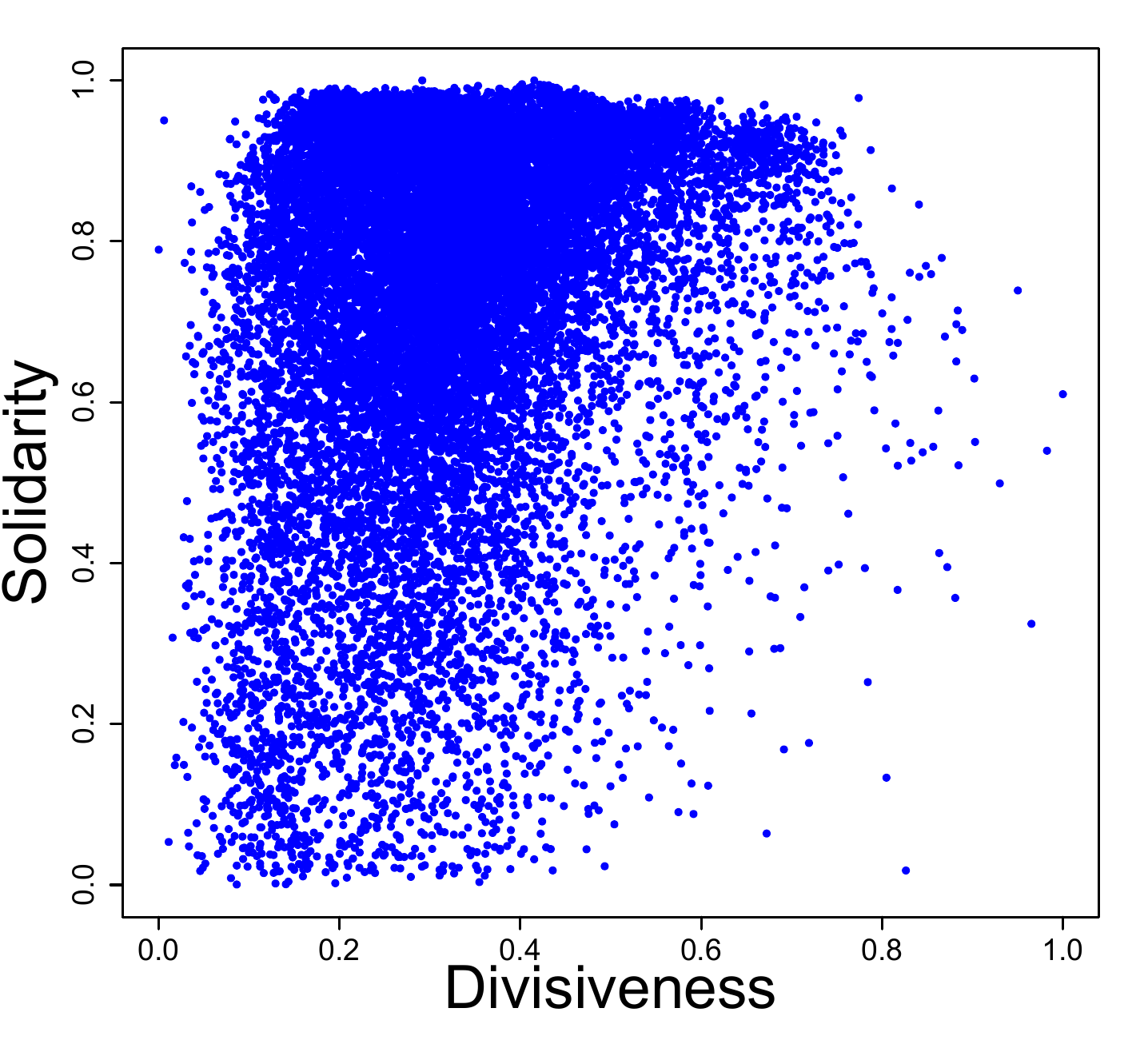}
}
\end{minipage}
\begin{minipage}[b]{0.5\linewidth}
\centerline{\it{Panel B: Divisiveness and Interaction}}
\centerline{
	\includegraphics[height=3in]{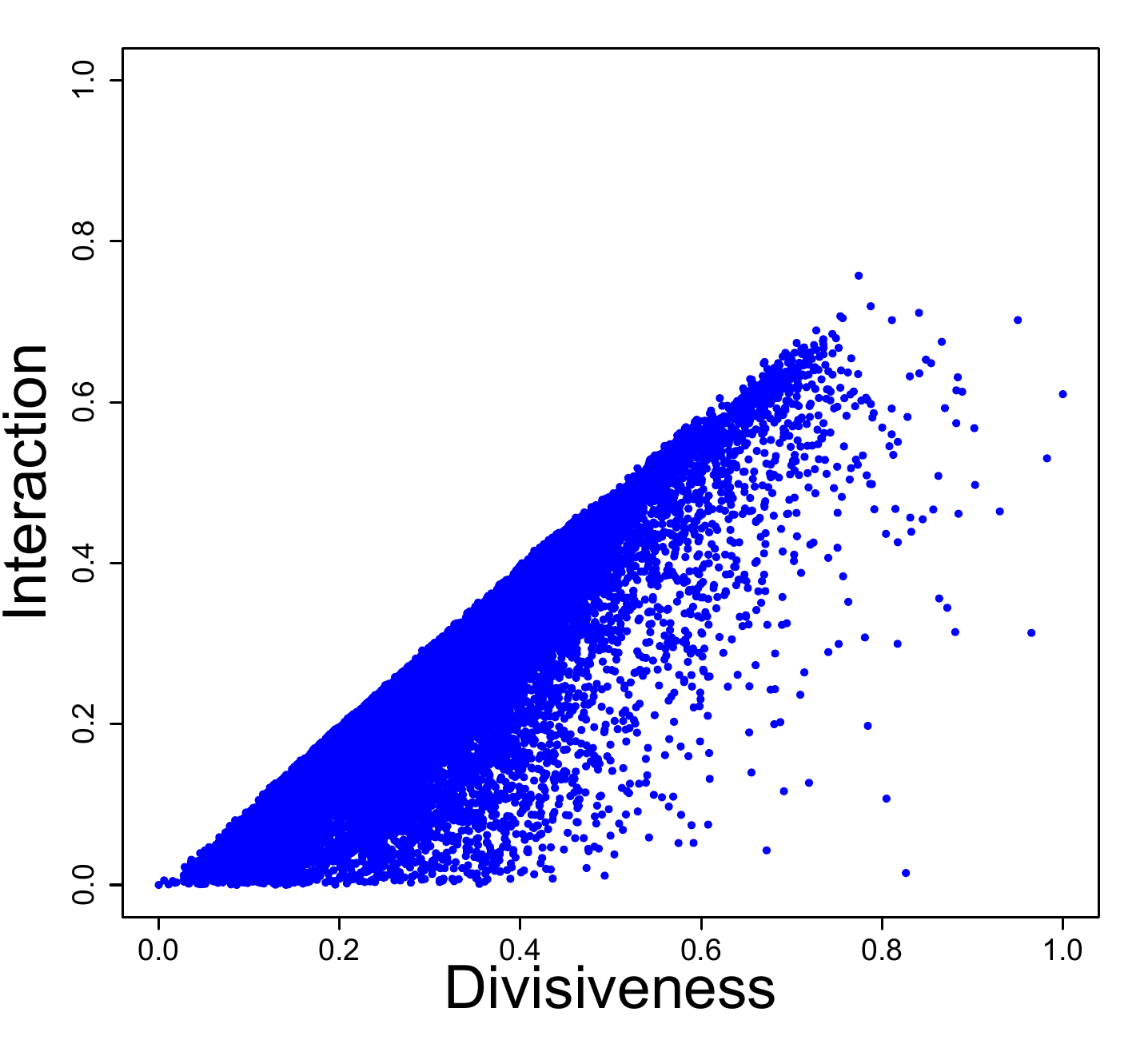}
}
\end{minipage}
\linebreak
\begin{minipage}[b]{0.5\linewidth}
\centerline{\it{Panel C: Solidarity and Interaction}}
\centerline{
	\includegraphics[height=3in]{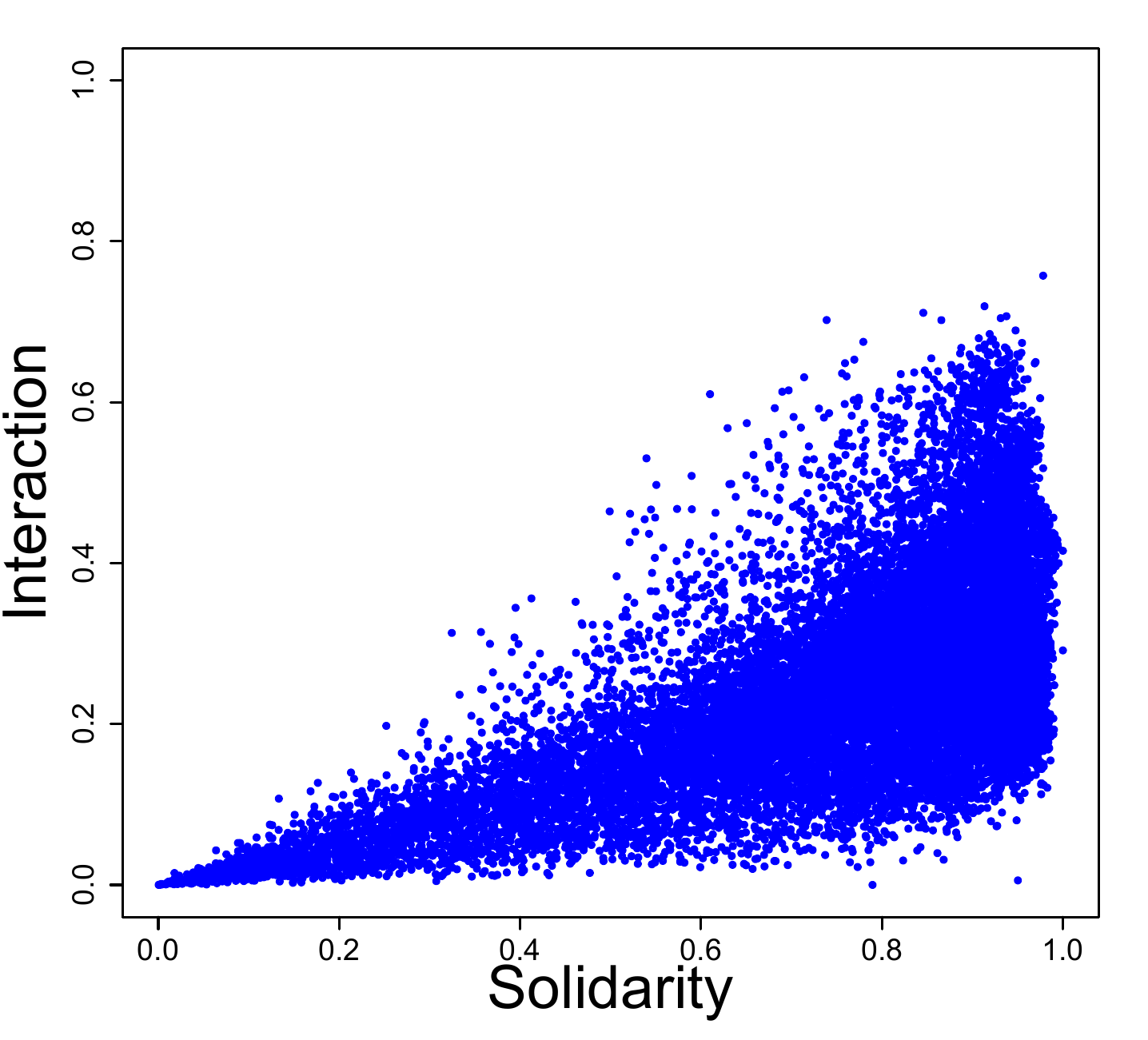}
}
\end{minipage}
\caption{[Color online] Scatter plots illustrating the relationship between divisiveness, solidarity, and their interaction for the House of Representatives.  All variables are scaled to lie in the interval $[0,1]$.\label{indlevscat}
}
\end{figure}

The plots and tables in this section supplement section \ref{comcent} of the main text.  Note in figure \ref{indlevscat} and table \ref{indlevcorr} that divisiveness, solidarity, and their interaction (divisiveness $\times$ solidarity) are positively correlated but clearly provide different pieces of information.


\subsubsection{Two-Dimensional LOESS Plots}

In addition to the LOESS regressions discussed in section \ref{comcent}, here we also examine the relationship between divisiveness, solidarity, and the maximum-modularity level of Congress.  As expected from the definitions of these quantities, we find that high-modularity Congresses are characterized by high levels of divisiveness and solidarity, as they are composed of communities that are highly structured and partisan.  When either divisiveness or solidarity values dip, however, then medium-modularity Congresses become more likely.  Instability in medium-modularity cases appears to be driven either by divisiveness without solidarity or vice versa.  Intuitively, a legislator who is divisive but not solidary holds highly divisive positions while nevertheless being assigned to a large community (i.e., low solidarity).  A legislator who is solidary but not divisive tends to side with his/her community on most issues but holds broadly popular positions on other issues.

\begin{figure}[ht]
\begin{minipage}[b]{0.5\linewidth}
\centerline{\it{Panel A: Divisiveness, Modularity, Reelection}}
\centerline{
	\includegraphics[height=3in]{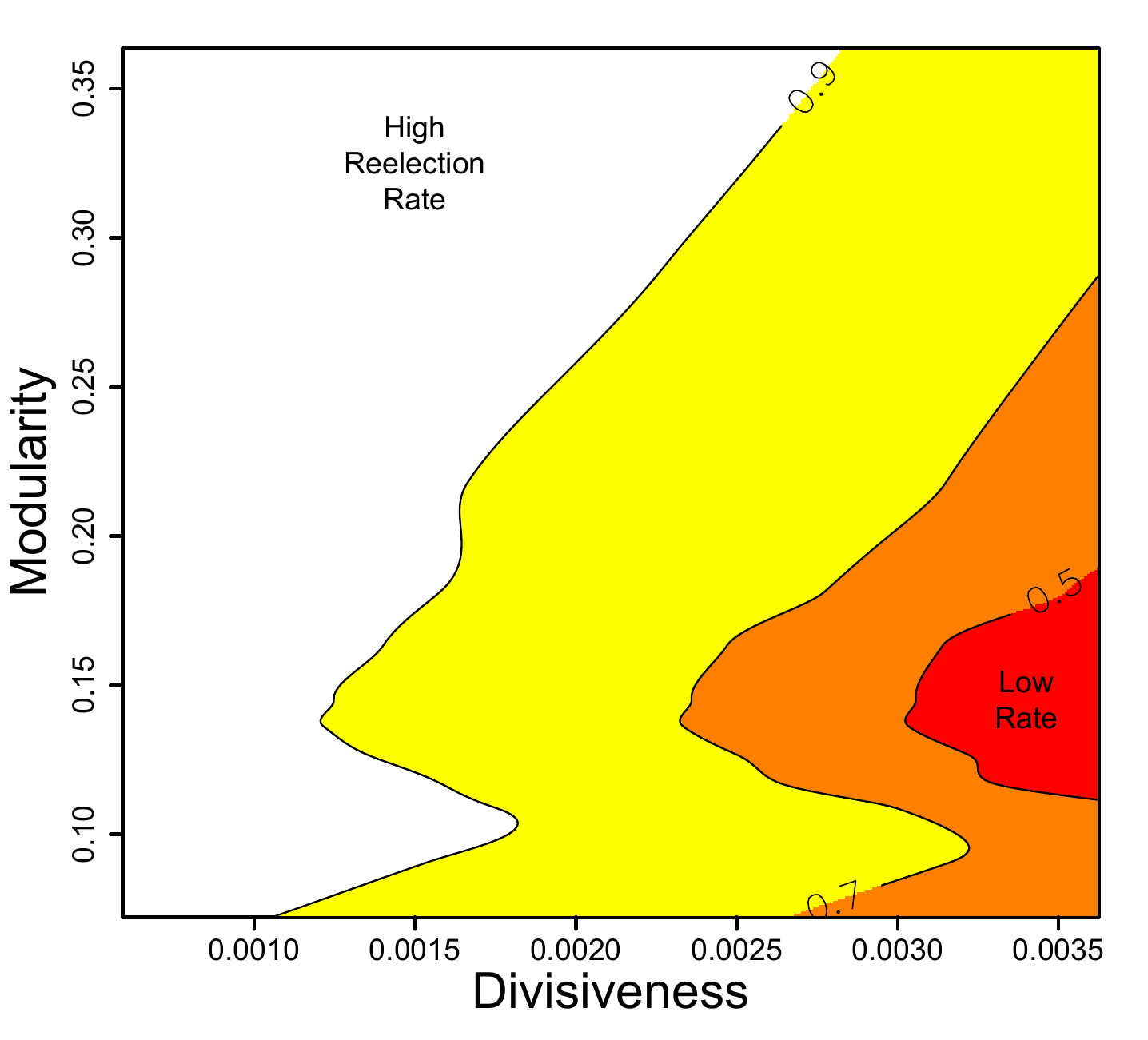}
}
\end{minipage}
\begin{minipage}[b]{0.5\linewidth}
\centerline{\it{Panel B: Divisiveness, Solidarity, Modularity}}
\centerline{
	\includegraphics[height=3in]{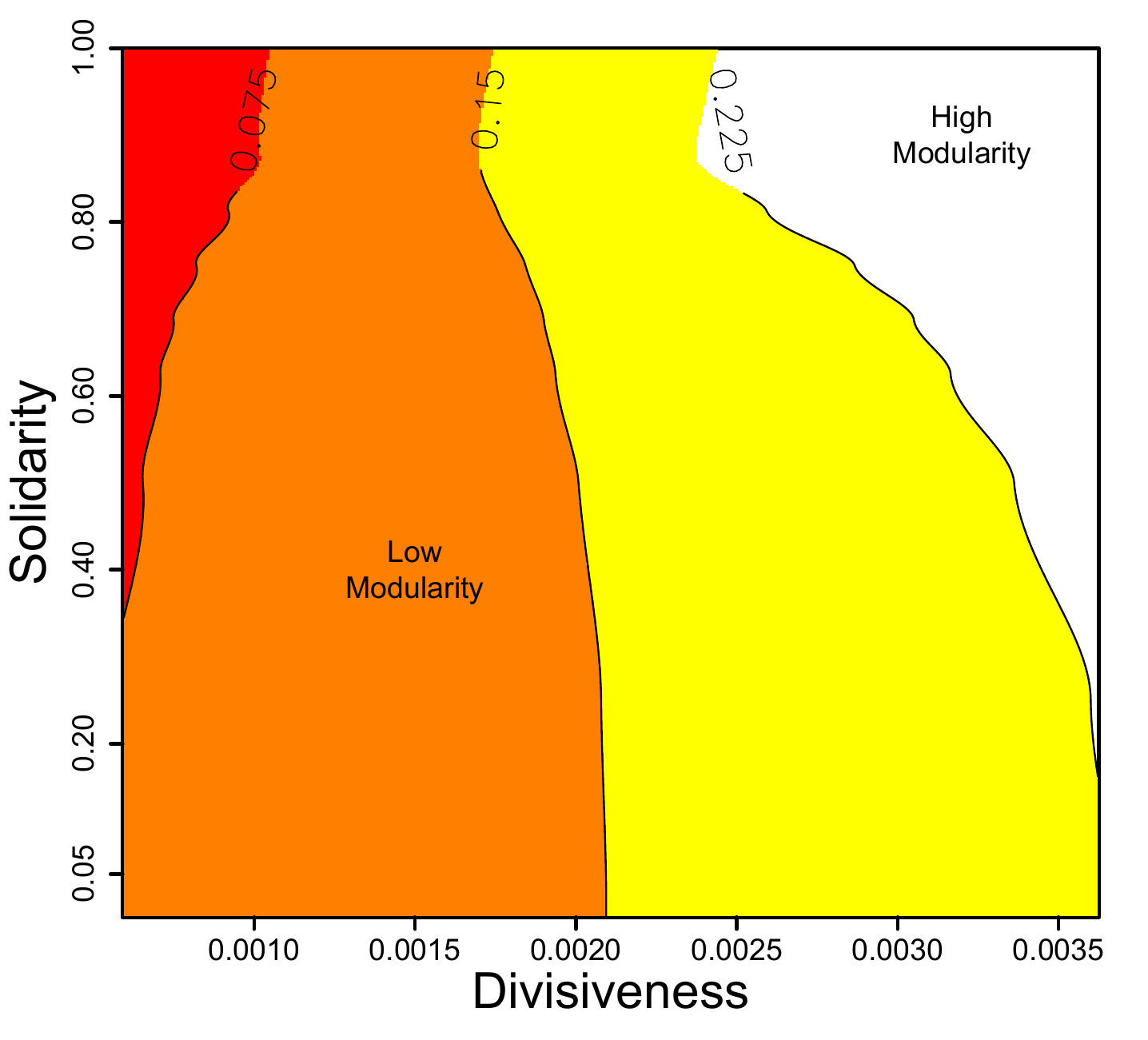}
}
\end{minipage}
\linebreak
\begin{minipage}[b]{0.5\linewidth}
\centerline{\it{Panel C: Divisiveness, Solidarity, Reelection}}
\centerline{
	\includegraphics[height=3in]{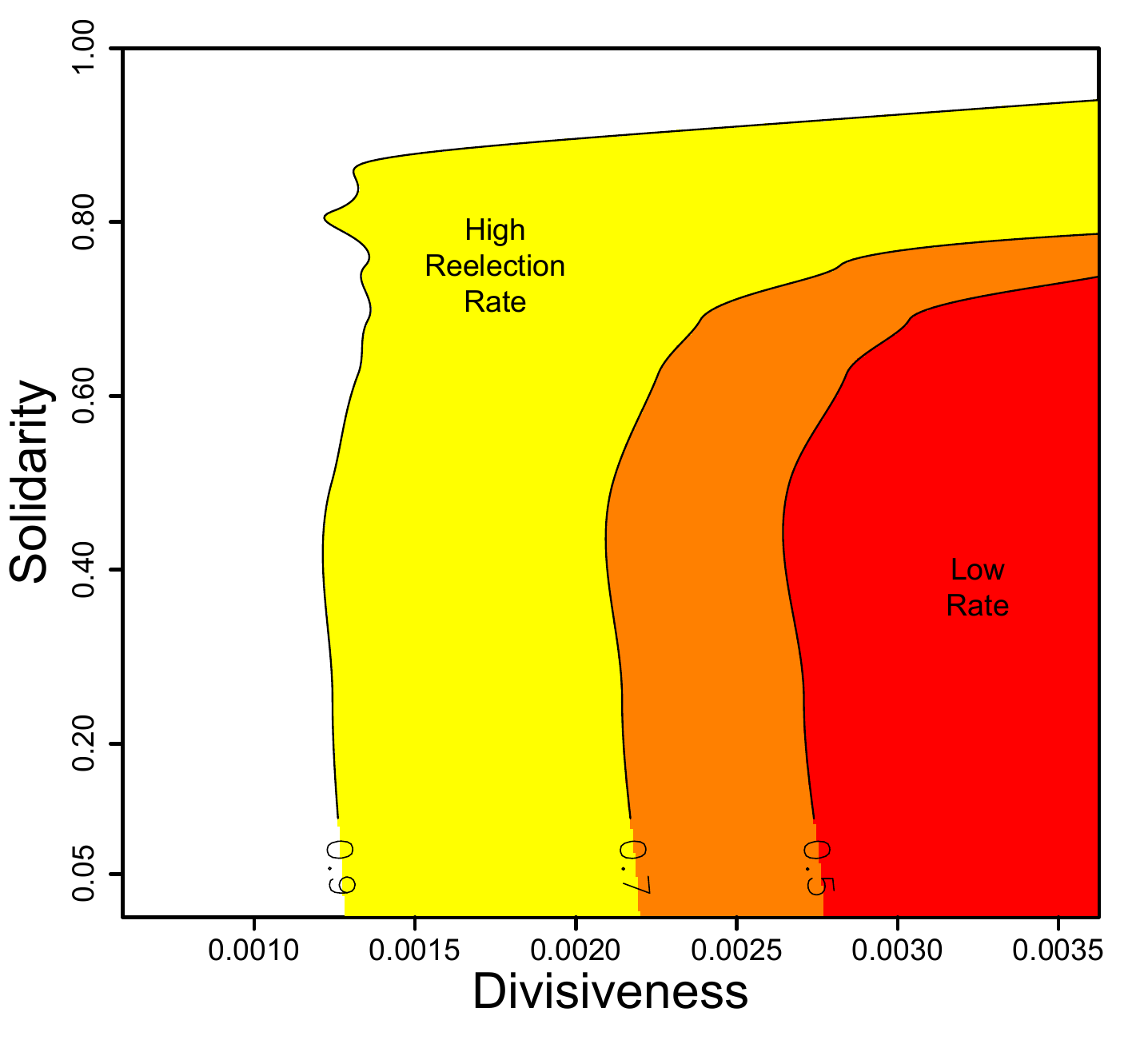}
}
\end{minipage}
\caption{[Color online] Two-Dimensional LOESS plots indicating the relationship between modularity, divisiveness, solidarity, and reelection rates in the House of Representatives.
\label{2dloess}}
\end{figure}

\begin{center}
$<$FIGURE \ref{2dloess} ABOUT HERE$>$
\end{center}

\subsubsection{Correlations and Scatter Plots}

\begin{center}
$<$TABLE \ref{indlevcorr} ABOUT HERE$>$
\end{center}

\begin{center}
$<$FIGURE \ref{indlevscat} ABOUT HERE$>$
\end{center}


\subsubsection{Divisiveness and Solidarity Summary Statistics}

In this section, we provide additional details on the divisiveness and solidarity of individual legislators.  Table \ref{divnotsol} gives summary statistics for legislators who are divisive (90th percentile or more) but not solidary (10th percentile or less), solidary but not divisive, both solidary and divisive, and neither solidary nor divisive.  In table \ref{dnsnames}, we give examples of legislators who fall into each of these four categories.

\begin{center}
$<$TABLE \ref{divnotsol} ABOUT HERE$>$
\end{center}

\begin{center}
$<$TABLE \ref{dnsnames} ABOUT HERE$>$
\end{center}



\pagebreak


\pagebreak



\pagebreak

\pagebreak

\pagebreak

\pagebreak
\begin{landscape}

\begin{table}[htbp]
\begin{center}
\scriptsize
\addtolength{\tabcolsep}{-5pt}
\begin{tabular}{|l|rllrll|rllrll|rllrll|rllrll|}
  	\multicolumn{25}{l}{\it Panel A: House Regressions} \\
  \hline
	 & \multicolumn{6}{c|}{\bf Maximum Modularity} & \multicolumn{6}{c|}{\bf 2-Dimensional W-NOM} & \multicolumn{6}{c|}{\bf Maximum Modularity (46:109)} & \multicolumn{6}{c|}{\bf 2-Dimensional DW-NOM (46:109)} \\
  \hline
	Majority Change & 0.556 & (0.51) &     & 0.748 & (0.63) &     & 0.729 & (0.51) &     & 1.125 & (0.64) & .   & 0.757 & (0.79) &     & 1.451 & (1.07) &     & 0.608 & (0.7) &     & 1.121 & (0.91) &     \\ 
	Max Modularity & 57.358 & (23.47) & *   & 61.168 & (25.17) & *   &  &  &  &  &  &  & 98.731 & (39.28) & *   & 160.85 & (60.68) & **  &  &  &  &  &  &  \\ 
	$[$Max Modularity$]^2$ & $-$123.153 & (60.84) & *   &$-$139.3 & (65.94) & *   &  &  &  &  &  &  & $-$213.359 & (97.54) & *   & $-$378.331 & (151.47) & *   &  &  &  &  &  &  \\ 
	$>2$ Communities & 1.006 & (0.55) & .   & 0.97 & (0.58) & .   &  &  &  &  &  &  & 2.089 & (0.94) & *   & 1.809 & (1.02) & .   &  &  &  &  &  &  \\ 
	W-NOM 2-Dim &  &  &  &  &  &  & $-$1.686 & (2.46) &     & $-$0.009 & (2.57) &     &  &  &  &  &  &  &  &  &  &  &  &  \\ 
	$[$W-NOM 2-Dim$]^2$ &  &  &  &  &  &  & 0.671 & (0.98) &     & $-$0.246 & (1.06) &     &  &  &  &  &  &  &  &  &  &  &  &  \\ 
	DW-NOM 2-Dim &  &  &  &  &  &  &  &  &  &  &  &  &  &  &  &  &  &  & $-$6.848 & (24.45) &     & 22.845 & (33.7) &     \\ 
	$[$DW-NOM 2-Dim$]^2$ &  &  &  &  &  &  &  &  &  &  &  &  &  &  &  &  &  &  & 6.028 & (17.76) &     & $-$16.736 & (25.05) &     \\ 
	Divided Government &  &  &  & $-$0.707 & (0.66) &     &  &  &  & $-$1.064 & (0.69) &     &  &  &  & $-$1.177 & (1.14) &     &  &  &  & $-$1.278 & (0.95) &     \\ 
	Midterm Congress &  &  &  & $-$0.511 & (0.53) &     &  &  &  & $-$0.339 & (0.52) &     &  &  &  & $-$1.594 & (1.14) &     &  &  &  & $-$0.872 & (0.78) &     \\ 
	2-year $\Delta$GDP &  &  &  & $-$0.004 & (0) &     &  &  &  & $-$0.005 & (0) &     &  &  &  & $-$0.008 & (0) & *   &  &  &  & $-$0.004 & (0) &     \\ 
	2-year $\Delta$CPI &  &  &  & 0.064 & (0.12) &     &  &  &  & 0.028 & (0.1) &     &  &  &  & 0.275 & (0.18) &     &  &  &  & 0.069 & (0.12) &     \\ 
	2-year $\Delta$Debt (\% GDP) &  &  &  &  &  &     &  &  &  &  &  &     &  &  &  &  &  &     &  &  &  &  &  &     \\ 
	(Intercept) & $-$7.168 & (2.18) & *** & $-$6.67 & (2.28) & **  & $-$0.38 & (1.49) &     & $-$0.247 & (1.53) &     & $-$11.717 & (3.83) & **  & $-$15.517 & (5.48) & **  & 0.282 & (7.99) &     & $-$7.804 & (10.75) &     \\ 
	   $\# Observations$ & \multicolumn{3}{|c}{108} & \multicolumn{3}{c|}{104} & \multicolumn{3}{c}{105} & \multicolumn{3}{c|}{102} & \multicolumn{3}{c}{63} & \multicolumn{3}{c|}{60} & \multicolumn{3}{c}{63} & \multicolumn{3}{c|}{60} \\
   \hline
   	\multicolumn{25}{c}{ }\\
  	\multicolumn{25}{l}{\it Panel B: Senate Regressions} \\
  \hline
	 & \multicolumn{6}{c|}{\bf Maximum Modularity} & \multicolumn{6}{c|}{\bf 2-Dimensional W-NOM} & \multicolumn{6}{c|}{\bf Maximum Modularity (46:109)} & \multicolumn{6}{c|}{\bf 2-Dimensional DW-NOM (46:109)} \\
  \hline
	Majority Change & 0.638 & (0.53) &     & 0.457 & (0.56) &     & 0.739 & (0.51) &     & 0.588 & (0.55) &     & 0.704 & (0.65) &     & 0.347 & (0.72) &     & 0.715 & (0.63) &     & 0.423 & (0.69) &     \\ 
	Max Modularity & 49.932 & (26.7) & .   & 47.287 & (28.02) & .   &  &  &  &  &  &  & 8.434 & (36.11) &     & 1.03 & (40.39) &     &  &  &  &  &  &  \\ 
	$[$Max Modularity$]^2$ & $-$150.818 & (80.47) & .   & $-$135.611 & (84.95) &     &  &  &  &  &  &  & $-$23.639 & (105.72) &     & 6.558 & (119.78) &     &  &  &  &  &  &  \\ 
	$>2$ Communities & $-$0.612 & (0.55) &     & $-$0.562 & (0.57) &     &  &  &  &  &  &  & 0.216 & (0.75) &     & 0.322 & (0.79) &     &  &  &  &  &  &  \\ 
	W-NOM 2-Dim &  &  &  &  &  &  & $-$10.598 & (8.58) &     & $-$13.625 & (9.5) &     &  &  &  &  &  &  &  &  &  &  &  &  \\ 
	$[$W-NOM 2-Dim$]^2$ &  &  &  &  &  &  & 4.053 & (3.59) &     & 5.321 & (3.96) &     &  &  &  &  &  &  &  &  &  &  &  &  \\ 
	DW-NOM 2-Dim &  &  &  &  &  &  &  &  &  &  &  &  &  &  &  &  &  &  & 14.732 & (25.19) &     & 11.761 & (33.74) &     \\ 
	$[$DW-NOM 2-Dim$]^2$ &  &  &  &  &  &  &  &  &  &  &  &  &  &  &  &  &  &  & $-$10.575 & (18.33) &     & $-$8.436 & (24.98) &     \\ 
	Divided Government &  &  &  & 0.199 & (0.53) &     &  &  &  & 0.317 & (0.53) &     &  &  &  & 0.038 & (0.7) &     &  &  &  & $-$0.034 & (0.73) &     \\ 
	Midterm Congress &  &  &  & 0.523 & (0.51) &     &  &  &  & 0.488 & (0.51) &     &  &  &  & 0.175 & (0.66) &     &  &  &  & 0.147 & (0.66) &     \\ 
	2-year $\Delta$GDP &  &  &  & $-$0.002 & (0) &     &  &  &  & $-$0.002 & (0) &     &  &  &  & $-$0.002 & (0) &     &  &  &  & $-$0.002 & (0) &     \\ 
	2-year $\Delta$CPI &  &  &  & 0.079 & (0.1) &     &  &  &  & 0.1 & (0.09) &     &  &  &  & 0.121 & (0.1) &     &  &  &  & 0.118 & (0.1) &     \\ 
	2-year $\Delta$Debt (\% GDP) &  &  &  &  &  &     &  &  &  &  &  &     &  &  &  &  &  &     &  &  &  &  &  &     \\ 
	(Intercept) & $-$4.63 & (2.18) & *   & $-$4.927 & (2.32) & *   & 5.325 & (5) &     & 6.789 & (5.58) &     & $-$2.075 & (2.94) &     & $-$1.764 & (3.29) &     & $-$6.183 & (8.29) &     & $-$5.146 & (10.81) &     \\
   	$\# Observations$ & \multicolumn{3}{|c}{108} & \multicolumn{3}{c|}{104} & \multicolumn{3}{c}{105} & \multicolumn{3}{c|}{102} & \multicolumn{3}{c}{63} & \multicolumn{3}{c|}{60} & \multicolumn{3}{c}{63} & \multicolumn{3}{c|}{60} \\
   \hline
   \hline
   	\multicolumn{25}{|c|}{Standard errors in parentheses.  Significance codes ($p<$): ï***Í 0.001, ï**Í 0.01, ï*Í 0.05, ï.Í 0.1}\\
   \hline
\end{tabular}
\caption{Logistic Regression Results: Panel A is the House; Panel B is the Senate. 
\label{modlongtab}}
\end{center}
\end{table}
\end{landscape}

\pagebreak
\begin{table}[htbp]
\begin{center}
\scriptsize
\addtolength{\tabcolsep}{-4pt}
\begin{tabular}{| l | rll  rll  rll  rll |}
 \multicolumn{13}{l}{\it Panel A: Fixed Effects}\\
\hline
								& \multicolumn{3}{c}{\bf (1) }	& \multicolumn{3}{c}{\bf (2) }	& \multicolumn{3}{c}{\bf (3) }	& \multicolumn{3}{c |}{\bf (4) }\\
\hline
	Divisiveness					& $-$4.737	& (0.267)	& ***		& $-$11.660	& (0.741)	& ***		& $-$12.728	& (0.761)	& ***		& $-$12.789	& (0.761)	& *** \\
	Solidarity						& 1.417	& (0.134)	& ***		& $-$1.700	& (0.346)	& ***		& $-$1.368	& (0.362)	& ***		& $-$1.353	& (0.362)	& *** \\
	Divisiveness$\times$Solidarity		&		&		&		& 10.030	& (0.992)	& ***		& 10.874	& (1.002)	& ***		& 10.990	& (1.003)	& *** \\
	Presidential Year				&		&		&		&		&		&		& $-$0.293	& (0.205)	&		& $-$0.299	& (0.202)	& \\
	Divided Government				&		&		&		&		&		&		& $-$0448	& (0.217)	& *		& $-$0.449	& (0.215)	& * \\
	 $|$Nominate (1st dim.)$ |$		&		&		&		&		&		&		& 0.967	& (0.247)	& ***		& 0.842	& (0.250)	& *** \\
	 Democrat						&		&		&		&		&		&		& $-$1.495	& (0.428)	& ***		& $-$1.723	& (0.469)	& *** \\
	 Republican					&		&		&		&		&		&		& $-$1.369	& (0.428)	& **		& $-$0.837	& (0.470)	& . \\
	 Seniority						&		&		&		&		&		&		& $-$0.062	& (0.009)	& ***		& $-$0.062	& (0.009)	& *** \\
	 Party Unity					&		&		&		&		&		&		& $-$0.010	& (0.003)	& ***		& $-$0.011	& (0.003)	& ***	\\
	 Victory Margin					&		&		&		&		&		&		& 0.038	& (0.002)	& ***		& 0.037	& (0.002)	& *** \\
	 Democrat$\times$Dem. Pres. Vote	&		&		&		&		&		&		&		&		&		& 0.005	& (0.004)	& \\
	 Republican$\times$Dem. Pres. Vote&		&		&		&		&		&		&		&		&		& $-$0.012	& (0.005)	& ** \\
	 (Intercept)					& 2.855	& (0.155)	& ***		& 4.932	& (0.267)	& ***		& 6.313	& (0.551)	& ***		& 6.395	& (0.552)	& *** \\
\hline
   	\multicolumn{13}{|c|}{Standard errors in parentheses.  Significance codes ($p<$): ï***Í 0.001, ï**Í 0.01, ï*Í 0.05, ï.Í 0.1.  $N= 16891$}\\
\hline
   	\multicolumn{13}{c}{ }\\
	\multicolumn{13}{l}{\it Panel B: Random Effects} \\
\hline
					& \multicolumn{3}{c}{\bf (1) }	& \multicolumn{3}{c}{\bf (2) }	& \multicolumn{3}{c}{\bf (3) }	& \multicolumn{3}{c |}{\bf (4) }\\
\hline
	Legislator (ICPSR)	& \multicolumn{3}{c}{0.933}	& \multicolumn{3}{c}{0.970}	& \multicolumn{3}{c}{0.633}	& \multicolumn{3}{c |} {0.633} \\
	(Number of Legislators $= 3867$)	& \multicolumn{3}{c}{(0.966)}	& \multicolumn{3}{c}{(0.985)}	& \multicolumn{3}{c}{(0.796)}	& \multicolumn{3}{c |}{(0.796)} \\	
	Congress (\#)		& \multicolumn{3}{c}{0.461}	& \multicolumn{3}{c}{0.430}	& \multicolumn{3}{c}{0.436}	& \multicolumn{3}{c |} {0.422} \\
	 (Number of Congresses $= 48$)	& \multicolumn{3}{c}{(0.679)}	& \multicolumn{3}{c}{(0.655)}	& \multicolumn{3}{c}{(0.661)}	& \multicolumn{3}{c |}{(0.650)} \\	 
\hline
 	\multicolumn{13}{|c|}{Variances reported with standard deviations in parentheses.}\\
\hline													
\end{tabular}
\caption{Mixed-Effects Logistic Regression Results for the 56th--103rd Houses.  The dependent variable is reelection to the House.  The key independent variables are divisiveness, solidarity, and their interaction.  Note that divisiveness and solidarity individually have a negative impact on electoral prospects but that the interaction has a positive impact.  This suggests that divisiveness might only be sustainable for Congressmen who are also strong members of a community. \label{indmelogit}}
\end{center}
\end{table}
\pagebreak


\pagebreak
\begin{table}[ht]
\begin{center}
\scriptsize
\addtolength{\tabcolsep}{-4pt}
\begin{tabular}{rlll}
\hline
	 \# & Method	& House	& Senate \\
	 \hline
	 1 & Leading-Eigenvector Spectral 1 \cite{newman2006,newman2006b} & 76 (76) & 49 (49) \\
	 2 & Leading-Eigenvector Spectral 2 \cite{newman2006,newman2006b} & 75 (1)& 48 (4) \\
	 3 & Leading-Eigenvector Spectral 3  \cite{newman2006,newman2006b} & 75 (1)& 45 (0) \\
	 4 & Two-Vector Bi-partitioning \cite{rich2009} & 81 (5)& 63 (9)\\
	 5 & Two-Vector Bi/Tri-partitioning \cite{rich2009} & 90 (9) & 69 (3)\\
	 6 & Louvain \cite{blondel2008} & 91 (14) & 88 (39) \\
	 7 & Simulated Annealing \cite{reich2006} & 83 (3) & 66 (5) \\
	 8& Walktrap \cite{pons2005} & 0 (0)& 0 (0) \\
	 9 & PAM Cluster Analysis \cite{kaufman1990} & 0 (0)& 0 (0) \\
	\hline
	  & Mean Modularity Interval & 0.0041 & 0.0166 \\
	  & Mean Identical to `Maximum-Modularity' Partition (minimum 1) & 5.8807 & 4.6239 \\
	 \hline
\end{tabular}
\caption{Summary Statistics for Community-Detection Heuristics.  In this table, we compare the partitions that we obtained using the eight modularity-optimization heuristics and the one that we obtained using a standard cluster-analysis technique.  Rows 1--9 give the number of Congresses (out of 109) for which each measure finds the `maximum-modularity' partition.  In parentheses, we report the number of Congresses for which we subsequently used the results of each method in our analyses.  Row 10 gives the mean modularity interval for the nine methods.  Row 11 gives the mean number of heuristics that find the `maximum-modularity' community partition.  These results suggest that community partitions in Congress are fairly robust to different heuristics for optimizing modularity.\label{heusums}}
\end{center}
\end{table}

\pagebreak
\begin{table}[ht]
\begin{center}
\scriptsize
\addtolength{\tabcolsep}{-4pt}
\begin{tabular}{rl p{1.5in} p{1.5in}}
	\hline
	 \# & Method	& House	& Senate \\
	 \hline
	 1 & Leading-Eigenvector Spectral 1 \cite{newman2006,newman2006b} & 3--13, 16, 23--26, 28, 30--36, 39, 41--43, 45--49, 51, 53--60, 62--72, 74, 76--78, 82--84, 87--89, 93, 94, 97--99, 100, 102, 104--109 & 2, 4, 7, 11--13, 16, 19, 22--24, 26--28, 32, 36, 43--45, 47, 48, 50--53, 55, 57--64, 66, 67, 82, 84, 87, 90, 93, 94, 103--109 \\
	 2 & Leading-Eigenvector Spectral 2 \cite{newman2006,newman2006b} & 79 & 1, 21, 33, 37 \\
	 3 & Leading-Eigenvector Spectral 3  \cite{newman2006,newman2006b} & 40 & none \\
	 4 & Two-Vector Bi-partitioning \cite{rich2009} & 20, 22, 44, 61, 80 & 15, 42, 56, 75, 86, 95, 96, 100, 102\\
	 5 & Two-Vector Bi/Tri-partitioning \cite{rich2009} & 2, 15, 17, 18, 21, 27, 37, 38, 91 & 29, 39, 65\\
	 6 & Louvain \cite{blondel2008} & 14, 29, 50, 52, 73, 75, 81, 85, 90, 92, 95, 96, 101, 103 & 3, 5, 6, 8, 10, 14, 17, 18, 20, 25, 30, 31, 34, 35, 38, 40, 41, 46, 49, 68-74, 77, 79--81, 83, 88, 89, 91, 92, 97, 97--101 \\
	 7 & Simulated Annealing \cite{reich2006} & 1, 19, 86 & 9, 54, 76, 78, 85 \\
	 8& Walktrap \cite{pons2005} & none & none \\
	 9 & PAM Cluster Analysis \cite{kaufman1990} & none & none \\
	\hline
\end{tabular}
\caption{This table lists the specific congresses for which each community-detection heuristic yielded the `maximum-modularity' partition.  These partitions were subsequently used for individual-level analyses in section \ref{comcent}.\label{heusums2}}
\end{center}
\end{table}

\pagebreak
\begin{landscape}
\begin{table}[ht]
\begin{center}
\scriptsize
\addtolength{\tabcolsep}{-4pt}

\begin{tabular}{| l	| rl rl rl |	 rl rl rl 	|	rl rl rl 	|	rl rl rl |}
  \multicolumn{25}{l}{\it Panel A: House}\\
  \hline
{\bf Congress} & \multicolumn{6}{|c|}{\bf Community 1 } & \multicolumn{6}{c|}{\bf Community 2 } & \multicolumn{6}{c|}{\bf Community 3 } & \multicolumn{6}{c|}{\bf Community 4 } \\
& \multicolumn{2}{|c}{\it Size} & \multicolumn{2}{c}{\it Divisiveness}  & \multicolumn{2}{c|}{\it Solidarity} & \multicolumn{2}{c}{\it Size} & \multicolumn{2}{c}{\it Divisiveness}  & \multicolumn{2}{c|}{\it Solidarity} & \multicolumn{2}{c}{\it Size} & \multicolumn{2}{c}{\it Divisiveness}  & \multicolumn{2}{c|}{\it Solidarity} & \multicolumn{2}{c}{\it Size} & \multicolumn{2}{c}{\it Divisiveness}  & \multicolumn{2}{c|}{\it Solidarity} \\
  \hline
2 & 33 & (46.5) & 0.01038 & (0.00246) & 0.75 & (0.214) & 32 & (45.1) & 0.01019 & (0.00244) & 0.79 & (0.216) & 5 & (7) & 0.00879 & (0.0013) & 0.93 & (0.057) & 1 & (1.4) & 0.00533 &  & 1 &  \\ 
  9 & 67 & (45.6) & 0.00445 & (0.00093) & 0.62 & (0.193) & 45 & (30.6) & 0.00499 & (0.00181) & 0.58 & (0.2) & 35 & (23.8) & 0.00547 & (0.00089) & 0.79 & (0.211) &  &  &  &  &  &  \\ 
  14 & 80 & (41) & 0.00418 & (0.00091) & 0.5 & (0.179) & 71 & (36.4) & 0.00445 & (0.0011) & 0.58 & (0.207) & 44 & (22.6) & 0.00429 & (0.00096) & 0.57 & (0.168) &  &  &  &  &  &  \\ 
  15 & 96 & (49.2) & 0.00416 & (0.00072) & 0.52 & (0.141) & 96 & (49.2) & 0.00439 & (0.00115) & 0.48 & (0.172) & 3 & (1.5) & 0.00349 & (0.00034) & 0.68 & (0.093) &  &  &  &  &  &  \\ 
  17 & 78 & (39.2) & 0.00382 & (0.00087) & 0.39 & (0.158) & 65 & (32.7) & 0.00374 & (0.00101) & 0.4 & (0.162) & 56 & (28.1) & 0.00404 & (0.00107) & 0.49 & (0.213) &  &  &  &  &  &  \\ 
  21 & 110 & (50) & 0.00333 & (0.00071) & 0.64 & (0.283) & 109 & (49.5) & 0.00317 & (0.00079) & 0.69 & (0.254) & 1 & (0.5) & 0.00222 &  & 1 &  &  &  &  &  &  &  \\ 
  32 & 100 & (41.8) & 0.00287 & (5e-04) & 0.57 & (0.189) & 87 & (36.4) & 0.00293 & (0.00075) & 0.57 & (0.214) & 52 & (21.8) & 0.00271 & (0.00044) & 0.65 & (0.194) &  &  &  &  &  &  \\ 
  52 & 160 & (47.1) & 0.00201 & (5e-04) & 0.6 & (0.214) & 91 & (26.8) & 0.00255 & (0.00055) & 0.78 & (0.159) & 89 & (26.2) & 0.00202 & (0.00047) & 0.61 & (0.158) &  &  &  &  &  &  \\ 
  81 & 172 & (38.7) & 0.00172 & (0.00033) & 0.79 & (0.159) & 169 & (38.1) & 0.00155 & (0.00037) & 0.82 & (0.181) & 103 & (23.2) & 0.00161 & (0.00028) & 0.79 & (0.095) &  &  &  &  &  &  \\ 
  85 & 194 & (43.7) & 0.00167 & (0.00044) & 0.66 & (0.177) & 180 & (40.5) & 0.00157 & (0.00028) & 0.71 & (0.216) & 70 & (15.8) & 0.00168 & (0.00022) & 0.79 & (0.118) &  &  &  &  &  &  \\ 
  86 & 211 & (47.6) & 0.00142 & (3e-04) & 0.75 & (0.183) & 161 & (36.3) & 0.00184 & (0.00036) & 0.75 & (0.193) & 71 & (16) & 0.00163 & (0.00032) & 0.78 & (0.165) &  &  &  &  &  &  \\ 
   \hline
   \multicolumn{25}{c}{}\\
   \multicolumn{25}{l}{\it Panel B: Senate}\\
   \hline
{\bf Congress} & \multicolumn{6}{|c|}{\bf Community 1 } & \multicolumn{6}{c|}{\bf Community 2 } & \multicolumn{6}{c|}{\bf Community 3 } & \multicolumn{6}{c|}{\bf Community 4 } \\
& \multicolumn{2}{|c}{\it Size} & \multicolumn{2}{c}{\it Divisiveness}  & \multicolumn{2}{c|}{\it Solidarity} & \multicolumn{2}{c}{\it Size} & \multicolumn{2}{c}{\it Divisiveness}  & \multicolumn{2}{c|}{\it Solidarity} & \multicolumn{2}{c}{\it Size} & \multicolumn{2}{c}{\it Divisiveness}  & \multicolumn{2}{c|}{\it Solidarity} & \multicolumn{2}{c}{\it Size} & \multicolumn{2}{c}{\it Divisiveness}  & \multicolumn{2}{c|}{\it Solidarity} \\
  \hline
1 & 10 & (34.5) & 0.02379 & (0.0026) & 0.78 & (0.161) & 10 & (34.5) & 0.02031 & (0.00412) & 0.67 & (0.123) & 9 & (31) & 0.02645 & (0.00712) & 0.71 & (0.176) &  &  &  &  &  &  \\ 
  2 & 14 & (45.2) & 0.02164 & (0.00356) & 0.82 & (0.117) & 10 & (32.3) & 0.02609 & (0.0064) & 0.83 & (0.133) & 7 & (22.6) & 0.02117 & (0.00453) & 0.92 & (0.096) &  &  &  &  &  &  \\ 
  4 & 15 & (34.9) & 0.01967 & (0.00546) & 0.81 & (0.14) & 15 & (34.9) & 0.02352 & (0.0053) & 0.79 & (0.188) & 13 & (30.2) & 0.02014 & (0.00504) & 0.79 & (0.182) &  &  &  &  &  &  \\ 
  5 & 20 & (45.5) & 0.0184 & (0.00694) & 0.64 & (0.213) & 15 & (34.1) & 0.02229 & (0.00495) & 0.76 & (0.188) & 9 & (20.5) & 0.01913 & (0.00624) & 0.77 & (0.151) &  &  &  &  &  &  \\ 
  8 & 18 & (40.9) & 0.01487 & (0.00701) & 0.58 & (0.208) & 13 & (29.5) & 0.0194 & (0.00372) & 0.76 & (0.285) & 13 & (29.5) & 0.01641 & (0.00606) & 0.72 & (0.129) &  &  &  &  &  &  \\ 
  9 & 16 & (43.2) & 0.01499 & (0.00421) & 0.8 & (0.21) & 14 & (37.8) & 0.01819 & (0.0055) & 0.69 & (0.276) & 7 & (18.9) & 0.01349 & (0.00712) & 0.67 & (0.166) &  &  &  &  &  &  \\ 
  11 & 18 & (40.9) & 0.01514 & (0.00447) & 0.6 & (0.245) & 13 & (29.5) & 0.01979 & (0.00448) & 0.74 & (0.201) & 13 & (29.5) & 0.01677 & (0.00743) & 0.68 & (0.144) &  &  &  &  &  &  \\ 
  14 & 14 & (31.8) & 0.01338 & (0.0039) & 0.7 & (0.11) & 13 & (29.5) & 0.01424 & (0.00352) & 0.76 & (0.18) & 12 & (27.3) & 0.01226 & (0.00549) & 0.66 & (0.178) & 5 & (11.4) & 0.01428 & (0.00472) & 0.82 & (0.087) \\ 
  15 & 19 & (41.3) & 0.0121 & (0.00405) & 0.54 & (0.246) & 17 & (37) & 0.0133 & (0.00397) & 0.63 & (0.185) & 10 & (21.7) & 0.01405 & (0.00783) & 0.83 & (0.069) &  &  &  &  &  &  \\ 
  17 & 19 & (36.5) & 0.01197 & (0.00493) & 0.65 & (0.207) & 17 & (32.7) & 0.01193 & (0.00253) & 0.61 & (0.177) & 15 & (28.8) & 0.01121 & (0.00339) & 0.63 & (0.172) & 1 & (1.9) & 0.01627 &  & 1 &  \\ 
  20 & 26 & (49.1) & 0.01287 & (0.0037) & 0.74 & (0.2) & 23 & (43.4) & 0.01305 & (0.00286) & 0.81 & (0.202) & 4 & (7.5) & 0.01225 & (0.00265) & 0.82 & (0.057) &  &  &  &  &  &  \\ 
  28 & 28 & (49.1) & 0.01355 & (0.0025) & 0.82 & (0.181) & 28 & (49.1) & 0.01137 & (0.00135) & 0.95 & (0.089) & 1 & (1.8) & 0.00258 &  & 1 &  &  &  &  &  &  &  \\ 
  30 & 36 & (50) & 0.01152 & (0.00403) & 0.53 & (0.197) & 26 & (36.1) & 0.01253 & (0.00372) & 0.66 & (0.169) & 10 & (13.9) & 0.0114 & (0.00203) & 0.69 & (0.234) &  &  &  &  &  &  \\ 
  31 & 29 & (41.4) & 0.01179 & (0.00282) & 0.65 & (0.242) & 27 & (38.6) & 0.01086 & (0.00215) & 0.71 & (0.212) & 14 & (20) & 0.0103 & (0.00326) & 0.65 & (0.175) &  &  &  &  &  &  \\ 
  32 & 29 & (39.7) & 0.0093 & (0.0025) & 0.64 & (0.222) & 25 & (34.2) & 0.00965 & (0.00401) & 0.56 & (0.241) & 19 & (26) & 0.00942 & (0.00234) & 0.63 & (0.146) &  &  &  &  &  &  \\ 
  33 & 24 & (34.3) & 0.00893 & (0.00207) & 0.62 & (0.178) & 24 & (34.3) & 0.00898 & (0.00431) & 0.54 & (0.192) & 22 & (31.4) & 0.00968 & (0.00255) & 0.59 & (0.195) &  &  &  &  &  &  \\ 
  40 & 33 & (47.8) & 0.00796 & (0.00122) & 0.68 & (0.164) & 21 & (30.4) & 0.01086 & (0.00334) & 0.78 & (0.21) & 15 & (21.7) & 0.00882 & (0.00662) & 0.64 & (0.252) &  &  &  &  &  &  \\ 
  41 & 38 & (47.5) & 0.00683 & (0.00283) & 0.56 & (0.192) & 22 & (27.5) & 0.00924 & (0.00338) & 0.71 & (0.274) & 20 & (25) & 0.00675 & (0.00303) & 0.64 & (0.203) &  &  &  &  &  &  \\ 
  43 & 28 & (35.4) & 0.01027 & (0.00228) & 0.8 & (0.17) & 27 & (34.2) & 0.00829 & (0.00258) & 0.81 & (0.166) & 24 & (30.4) & 0.00875 & (0.00199) & 0.81 & (0.188) &  &  &  &  &  &  \\ 
  54 & 42 & (46.7) & 0.00716 & (0.00128) & 0.68 & (0.174) & 39 & (43.3) & 0.00678 & (0.00112) & 0.77 & (0.183) & 9 & (10) & 0.00616 & (0.00097) & 0.83 & (0.125) &  &  &  &  &  &  \\ 
  70 & 47 & (46.1) & 0.00585 & (0.00171) & 0.71 & (0.194) & 39 & (38.2) & 0.00713 & (0.00158) & 0.75 & (0.271) & 16 & (15.7) & 0.007 & (0.00145) & 0.82 & (0.194) &  &  &  &  &  &  \\ 
  71 & 54 & (49.5) & 0.00661 & (0.00132) & 0.78 & (0.172) & 54 & (49.5) & 0.00707 & (0.00201) & 0.74 & (0.267) & 1 & (0.9) & 0.00431 &  & 1 &  &  &  &  &  &  &  \\ 
  72 & 38 & (36.9) & 0.00531 & (0.00142) & 0.64 & (0.201) & 37 & (35.9) & 0.00613 & (0.0013) & 0.8 & (0.215) & 28 & (27.2) & 0.00646 & (0.00189) & 0.72 & (0.201) &  &  &  &  &  &  \\ 
  74 & 37 & (37) & 0.00497 & (0.00154) & 0.75 & (0.181) & 33 & (33) & 0.0068 & (0.00169) & 0.76 & (0.193) & 30 & (30) & 0.00615 & (0.00145) & 0.72 & (0.164) &  &  &  &  &  &  \\ 
  78 & 48 & (46.2) & 0.00628 & (0.00185) & 0.66 & (0.246) & 46 & (44.2) & 0.00648 & (0.00156) & 0.69 & (0.254) & 10 & (9.6) & 0.00595 & (0.00121) & 0.85 & (0.181) &  &  &  &  &  &  \\ 
  83 & 53 & (48.2) & 0.00674 & (0.00191) & 0.68 & (0.262) & 51 & (46.4) & 0.00676 & (0.00196) & 0.73 & (0.17) & 6 & (5.5) & 0.00619 & (0.00242) & 0.69 & (0.141) &  &  &  &  &  &  \\ 
  85 & 44 & (43.6) & 0.00613 & (0.00126) & 0.77 & (0.215) & 41 & (40.6) & 0.00619 & (0.00116) & 0.78 & (0.215) & 16 & (15.8) & 0.00676 & (0.00239) & 0.83 & (0.154) &  &  &  &  &  &  \\ 
  88 & 50 & (49) & 0.00476 & (0.00117) & 0.85 & (0.181) & 31 & (30.4) & 0.00604 & (0.00143) & 0.83 & (0.16) & 21 & (20.6) & 0.00717 & (0.00113) & 0.92 & (0.101) &  &  &  &  &  &  \\ 
  92 & 50 & (49) & 0.00549 & (0.00144) & 0.76 & (0.234) & 46 & (45.1) & 0.00533 & (0.00146) & 0.85 & (0.216) & 6 & (5.9) & 0.0036 & (0.00044) & 0.98 & (0.014) &  &  &  &  &  &  \\ 
  98 & 49 & (48.5) & 0.0046 & (0.00114) & 0.83 & (0.221) & 47 & (46.5) & 0.00479 & (0.00124) & 0.8 & (0.272) & 5 & (5) & 0.00216 & (0.00118) & 0.92 & (0.106) &  &  &  &  &  &  \\ 
  99 & 50 & (49.5) & 0.00473 & (0.00132) & 0.87 & (0.186) & 48 & (47.5) & 0.00488 & (0.00152) & 0.85 & (0.232) & 3 & (3) & 0.00268 & (0.00085) & 0.86 & (0.162) &  &  &  &  &  &  \\ 
   \hline
\end{tabular}
\caption{Descriptive statistics for selected Houses and Senates with three or more communities. This table shows Houses and Senates in which the third-largest community is greater than or equal to the difference between the sizes of the largest and second-largest communities.  We report the size (number of legislators) of each community with the percentage of total legislators in parentheses).  We also report mean divisiveness and mean solidarity with standard deviations in parentheses. \label{3comstats}}
\end{center}
\end{table}
\end{landscape}


\pagebreak
\begin{table}[ht]
\begin{center}
\scriptsize
\addtolength{\tabcolsep}{-4pt}

\begin{tabular}{| rrll |}
  \multicolumn{4}{l}{\it Panel A: House}\\
  \hline
	{\bf Year} & {\bf Congress} & {\bf Old} & {\bf New} \\ 
  \hline
	1792 & 3 & Fed & Dem-Rep \\ 
	1794 & 4 & Dem-Rep & Fed \\ 
	1800 & 7 & Fed & Dem-Rep \\ 
	1824 & 19 & Dem-Rep & Adams \\ 
	1826 & 20 & Adams & Jackson \\ 
	1836 & 25 & Jackson & Dem \\ 
	1840 & 27 & Dem & Whig \\ 
	1842 & 28 & Whig & Dem \\ 
	1846 & 30 & Dem & Whig \\ 
	1848 & 31 & Whig & Dem \\ 
	1854 & 34 & Dem & Rep \\ 
	1856 & 35 & Rep & Dem \\ 
	1858 & 36 & Dem & Rep \\ 
	1874 & 44 & Rep & Dem \\ 
	1880 & 47 & Dem & Rep \\ 
	1882 & 48 & Rep & Dem \\ 
	1888 & 51 & Dem & Rep \\ 
	1890 & 52 & Rep & Dem \\ 
	1894 & 54 & Dem & Rep \\ 
	1910 & 62 & Rep & Dem \\ 
	1918 & 66 & Dem & Rep \\ 
	1930 & 72 & Rep & Dem \\ 
	1946 & 80 & Dem & Rep \\ 
	1948 & 81 & Rep & Dem \\ 
	1952 & 83 & Dem & Rep \\ 
	1954 & 84 & Rep & Dem \\ 
	1994 & 104 & Dem & Rep \\ 
   \hline
   	\multicolumn{4}{c}{ } \\
	 \multicolumn{4}{l}{\it Panel B: Senate}\\
  \hline
	{\bf Year} & {\bf Congress} & {\bf Old} & {\bf New} \\ 
  \hline
	1800 & 7 & Fed & Dem-Rep \\ 
	1824 & 19 & Dem-Rep & Adams \\ 
	1826 & 20 & Adams & Jackson \\ 
	1832 & 23 & Dem & Anti-Jackson \\ 
	1834 & 24 & Anti-Jackson & Jackson \\ 
	1836 & 25 & Jackson & Dem \\ 
	1840 & 27 & Dem & Whig \\ 
	1844 & 29 & Whig & Dem \\ 
	1860 & 37 & Dem & Rep \\ 
	1878 & 46 & Rep & Dem \\ 
	1880 & 47 & Dem & Rep \\ 
	1892 & 53 & Rep & Dem \\ 
	1894 & 54 & Dem & Rep \\ 
	1912 & 63 & Rep & Dem \\ 
	1914 & 64 & Dem & Rep \\ 
	1918 & 66 & Rep & Dem \\ 
	1932 & 73 & Dem & Rep \\ 
	1946 & 80 & Rep & Dem \\ 
	1948 & 81 & Dem & Rep \\ 
	1952 & 83 & Rep & Dem \\ 
	1954 & 84 & Dem & Rep \\ 
	1980 & 97 & Rep & Dem \\ 
	1986 & 100 & Dem & Rep \\ 
	1994 & 104 & Rep & Dem \\ 
	2000 & 107 & Dem & Rep \\ 
	2002 & 108 & Rep & Dem \\ 
   \hline
\end{tabular}
\caption{Majority Party Switches in the U.S. Congress (1788--2002). \label{partyswitch}}
\end{center}
\end{table}


\pagebreak
\begin{table}[htbp]
\begin{center}
\scriptsize
\addtolength{\tabcolsep}{-4pt}
\begin{tabular}{l|r|r|r|}
				& Divisiveness	& Solidarity	& Interaction \\
	Divisiveness	& 1			&			& \\
	Solidarity		& 0.253		& 1			& \\
	Interaction	& 0.866		& 0.653		& 1 \\

\end{tabular}
\caption{Pearson correlations between Divisiveness, Solidarity, and their Interaction \label{indlevcorr}}
\end{center}
\end{table}

\pagebreak
\begin{landscape}
\begin{table}[ht]
\begin{center}
\scriptsize
\addtolength{\tabcolsep}{-4pt}
\begin{tabular}{l | rrrrr rl rl rl rl r}
  \hline
 						& $N$  & Unique $N$ 	& \% Dem & \%Rep 	& \% In Majority 	& \multicolumn{2}{l}{Mean Party Unity (s.d.)} &  \multicolumn{2}{l}{Mean Seniority (s.d.)} & \multicolumn{2}{l}{Mean Vict. Margin (s.d.)} &  \multicolumn{2}{l}{Mean Extremity (s.d.)} & \% Reelected \\ 
  \hline
  	Divisive not Solidary 	& 61 		& 58 			& 62.30 	& 37.70 	& 70.49 		& 68.23 & (14.35) 					& 4.20 & (2.87) 						& 35.60 & (31.30) 						& 0.30 & (0.18) & 49.18 \\ 
	Solidary not Divisive 	& 48 		& 40 			& 100.00 	& 0.00 	& 100.00 		& 86.06 & (8.45) 					& 4.54 & (3.65) 						& 29.90 & (24.20) 						& 0.30 & (0.09) & 91.67 \\ 
  	Solidary and Divisive 	& 192 	& 161 		& 50.00 	& 50.00 	& 19.79 		& 85.86 & (30.23) 					& 4.04 & (2.71) 						& 31.50 & (32.50) 						& 0.45 & (0.11) & 94.79 \\ 
  	Neither Div. nor Sol.		& 518 	& 232 		& 64.29 	& 35.71 	& 64.48 		& 57.01 & (10.25) 					& 4.85 & (3.69) 						& 39.00 & (29.00) 						& 0.12 & (0.09) & 93.24 \\ 
	All Legislators 			& 17776 	& 4037 		& 55.17 	& 44.42 	& 58.93 		& 81.39 & (16.77) 					& 4.50 & (3.50) 						& 35.80 & (31.10) 						& 0.33 & (0.17) & 88.33 \\ 
   \hline
\end{tabular}
\caption{Summary statistics for legislators who are divisive but not solidary, solidary but not divisive, solidary and divisive, and neither solidary nor divisive.  $N$ is the number of legislators in the category, and `unique $N$' is the number of unique legislators.  Also given are the percentage of Democrats and Republicans in each category, the percentage who are in the majority party, mean party unity, mean seniority (number of congresses served), mean ideological extremity (absolute value of 1st Dimension DW-NOM score), and the percent who were elected to the following House.\label{divnotsol}}
\end{center}
\end{table}
\end{landscape}

\pagebreak
\begin{table}[ht]
\begin{center}
\scriptsize
\addtolength{\tabcolsep}{-4pt}
\begin{tabular}{ rlllllll }
  \hline
 Congress & Name & State & District 		& Party & Divisiveness & Solidarity & Interaction \\ 
  \hline
  \hline 
  	\multicolumn{8}{c}{{\it Divisive but not Solidary} (10 least solidary listed)} \\
  	63 &  LOFT  G.W. 	&  NY	& 13 		& Dem & 0.8264 & 0.0179 & 0.0148 \\ 
  	77 &  WHIITEN    	&  MS	&  2 		& Dem & 0.6724 & 0.0639 & 0.0429 \\ 
  	66 &  SULLIVAN   	&  NY 	& 13 		& Dem & 0.5915 & 0.0881 & 0.0521  \\ 
  	61 &  BROUSSARD	&  LA 	&  3 		& Dem & 0.5748 & 0.0905 & 0.0520  \\ 
  	82 &  IKARD      	&  TX 	& 13 		& Dem & 0.5422 & 0.1087 & 0.0589 \\ 
  	75 &  CONNERY    	&  MA 	&  7 		& Dem & 0.6076 & 0.1233 & 0.0749  \\ 
  	71 &  OCONNELL   	&  RI 	&  3 		& Dem & 0.5892 & 0.1260 & 0.0742  \\ 
  	89 &  THOMPSON   	&  LA 	&  7 		& Dem & 0.5133 & 0.1331 & 0.0683 \\ 
  	59 &  ADAMS  H.C 	&  WI 	&  2 		& Rep & 0.8049 & 0.1333 & 0.1073  \\ 
  	59 &  PRINCE     	&  IL 		& 10 		& Rep & 0.5092 & 0.1496 & 0.0762 \\ 
     \hline 
       	\multicolumn{8}{c}{{\it Solidary but not Divisive} (10 least divisive listed)} \\
	95 	&  LUKEN      		&  OH    	&  1 		& Dem & 0.0059 & 0.9502 & 0.0056 \\ 
  	101 	&  CARR       		&  MI 	&  6 		& Dem & 0.0846 & 0.9488 & 0.0803 \\ 
  	101 	&  CARPER  T 		&  DE 	&  1 		& Dem & 0.1106 & 0.9531 & 0.1054 \\ 
  	97 	&  SMITH  N.  		&  IA    	&  5 		& Dem & 0.1152 & 0.9518 & 0.1097 \\ 
  	101	&  ENGLISH    		&  OK	&  6 		& Dem & 0.1154 & 0.9761 & 0.1127 \\ 
  	94 	&  VAN DEERLI 	&  CA 	& 37 		& Dem & 0.1193 & 0.9577 & 0.1143 \\ 
  	95	&  SPELLMAN   	& MD	&  5 		& Dem & 0.1204 & 0.9462 & 0.1139 \\ 
  	101 	&  SARPALIUS  	& TX   	& 13 		& Dem & 0.1229 & 0.9828 & 0.1208 \\ 
  	98 	&  MCNULTY  J 	& AZ		&  5 		& Dem & 0.1270 & 0.9788 & 0.1243 \\ 
  	100 	&  CARR       		& MI		&  6 		& Dem & 0.1279 & 0.9759 & 0.1248 \\ 

     \hline 
       	\multicolumn{8}{c}{{\it Solidary and Divisive} (10 highest interactions listed)} \\
	67 &  KITCHIN  C 		&  NC &  2 		& Dem & 0.7742 & 0.9782 & 0.7573 \\ 
	56 &  ELLIOTT    		&  SC &  7 		& Dem & 0.7272 & 0.9477 & 0.6892 \\ 
  	59 &  HOWARD  W. 		&  GA &  8 		& Dem & 0.7055 & 0.9549 & 0.6737 \\ 
  	58 &  POU  E.W.  		&  NC &  4 		& Dem & 0.6931 & 0.9541 & 0.6613 \\ 
  	58 &  RANDELL  C 		&  TX   &  5 		& Dem & 0.6944 & 0.9495 & 0.6594 \\ 
  	59 &  MCCLAIN    		&  MS &  6 		& Dem & 0.6932 & 0.9504 & 0.6588 \\ 
  	60 &  SMITH  W.R 		&  TX   & 16 		& Dem & 0.6901 & 0.9514 & 0.6566 \\ 
  	59 &  CANDLER  E 		&  MS &  1 		& Dem & 0.6701 & 0.9699 & 0.6499 \\ 
  	59 &  ROBINSON   		&  AR &  6 		& Dem & 0.6851 & 0.9466 & 0.6485 \\ 
  	58 &  WALLACE  R 		&  AR &  7 		& Dem & 0.6691 & 0.9689 & 0.6482 \\ 

     \hline 
       	\multicolumn{8}{c}{{\it Neither Solidary nor Divisive} (10 lowest interactions listed)}\\
	98 &  RINALDO W  		&  NJ & 12 		& Rep & 0.0863 & 0.0005 & 0.0000 \\ 
  	95 &  RINALDO W  		&  NJ & 12 		& Rep & 0.1406 & 0.0008 & 0.0001 \\ 
  	97 &  WHITLEY    		&  NC &  3 		& Dem & 0.1292 & 0.0017 & 0.0002 \\ 
  	102 &  RAY  R     		&  GA &  3 		& Dem & 0.1435 & 0.00416 & 0.0006 \\ 
  	94 &  DERRICK    		&  SC &  3 		& Dem & 0.0110 & 0.0533 & 0.0006 \\ 
  	93 &  STUCKEY    		&  GA &  8 		& Dem & 0.0799 & 0.0084 & 0.0007 \\ 
  	98 &  BENNETT  C 		&  FL &  2 			& Dem & 0.0459 & 0.017308 & 0.0008 \\ 
  	96 &  NELSON  C  		&  FL &  9 			& Dem & 0.0488 & 0.0213 & 0.0010 \\ 
  	95 &  GIBBONS    		&  FL & 10 		& Dem & 0.0515 & 0.0265 & 0.0014 \\ 
  	97 &  BENNETT  C 		&  FL &  2 			& Dem & 0.0430 & 0.0369 & 0.0016 \\ 
   \hline
\end{tabular}
\caption{Examples of House members who are divisive but not solidary, solidary but not divisive, solidary and divisive, and neither solidary nor divisive. 
\label{dnsnames}}
\end{center}
\end{table}

\pagebreak

\bibliographystyle{apsr}
\bibliography{modularity}

\end{document}